\documentclass[aps,pre,twocolumn,groupedaddress,floatfix,longbibliography,superscriptaddress]{revtex4-1}
\usepackage{amsmath}
\usepackage{amssymb}
\usepackage{amsfonts}
\usepackage{graphicx}
\usepackage{bbold}
\usepackage{bm}
\usepackage{bm}
\usepackage{cancel}
\usepackage{times,float}
\usepackage{graphicx}
\usepackage[usenames,dvipsnames,svgnames]{xcolor}
\usepackage{hyperref}
\hypersetup{colorlinks=true, linkcolor=Blue, citecolor=PineGreen,urlcolor=cyan}
\usepackage{multirow}

\usepackage{wrapfig}
\usepackage{mathtools,slashed}
\usepackage{braket}
\usepackage{caption}

\usepackage{siunitx}
\usepackage{ulem}
\usepackage{multirow}

\usepackage{pdfpages} 
\usepackage{pgffor} 

\def\supplementfilename{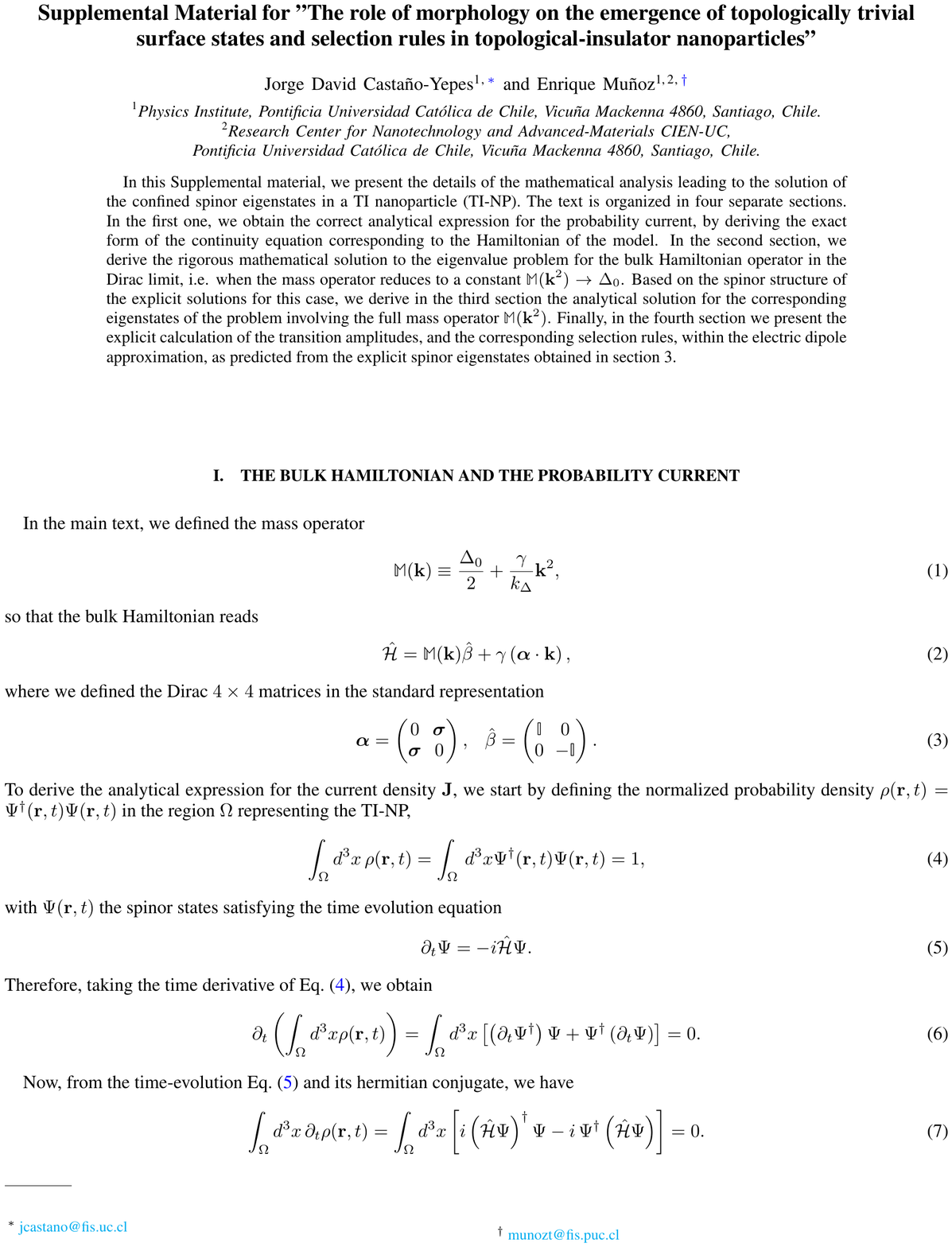}

\makeatletter
\AtBeginDocument{\let\LS@rot\@undefined}
\makeatother

\pdfximage{\supplementfilename}
\def\numbersupplementpages{\the\pdflastximagepages}

\newif\ifarXiv
\arXivtrue 

\newcommand{\bea}{\begin{eqnarray}}
\newcommand{\eea}{\end{eqnarray}}
\newcommand{\nn}{\nonumber}

\newcommand{\ii}{\mathrm{i}}
\newcommand{\Ham}{\hat{\mathcal{H}}}

\begin{document}
\title{The role of morphology on the emergence of topologically trivial surface states and selection rules in topological-insulator nano-particles}
\author{Jorge David Casta\~no-Yepes}
\email{jcastano@fis.uc.cl}
\affiliation{Physics Institute, Pontificia Universidad Católica de Chile, Vicuña Mackenna 4860, Santiago, Chile.}
\author{Enrique Muñoz}
\email{munozt@fis.puc.cl}
\address{Physics Institute, Pontificia Universidad Católica de Chile, Vicuña Mackenna 4860, Santiago, Chile.}
\address{Research Center for Nanotechnology and Advanced-Materials CIEN-UC, Pontificia Universidad Católica de Chile, Vicuña Mackenna 4860, Santiago, Chile.}


\begin{abstract}
Confined electronic states and optical transitions in 3D topological insulator nanoparticles have been studied in the literature, assuming idealized geometries such as spheres or infinitely long cylinders, that allow to obtain analytical solutions to the corresponding eigenvalue equation within such geometries. In contrast, in this article we consider triangular-shaped nanoplates as a more realistic approximation to the experimentally observed morphologies of topological insulator nanoparticles. In this particular geometry, we obtain analytical expressions for the confined eigenstates and the corresponding energy spectrum. Moreover, by a spatial representation of the probability density distribution of these states, we further identify the conditions leading to the emergence of topologically trivial surface states as a result of geometric confinement. Finally, we also study the optical transitions and the corresponding selection rules imposed by the nanoparticle size and morphology.
\end{abstract}
\maketitle

\section{Introduction}
Topological insulators (TIs) are materials presenting a large bulk band-gap, in combination with gapless topologically-protected~\cite{Chiu_16} edge (in 2D TIs) or surface (in 3D TIs) states with a nearly linear (Dirac-like) dispersion, as confirmed by angle-resolved photoemission spectroscopy (ARPES)~\cite{Hasan_10,Qi_11}. Moreover, these edge/surface non-trivial topological states exhibit a remarkable spin-momentum interlocking (chirality) property~~\cite{Roushan_09} that makes them exciting candidates for technological applications in quantum information, spintronics and energy harvesting in thermoelectric devices~\cite{Xu_17}.
The basic concept of a 2D TI was first proposed theoretically~\cite{Bernevig_06} and later discovered experimentally~\cite{Konig_07} in HgTe/CdTe heterostructures displaying the quantum spin Hall Effect.
On the other hand, the theoretical prediction ~\cite{Fu_Kane_Mele_07} of 3D TIs was rapidly followed by their experimental discovery in the Bi$_{x}$Sb$_{1-x}$ compounds~\cite{Hsieh_08}, and since then it has triggered an active research for other materials~\cite{Xia_09} that provide actual realizations of the concept. Important examples for their relatively large band-gap~\cite{Zhang_09,Xia_09} are the metal dichalcogenides Bi$_{2}$Se$_{3}$, Bi$_{2}$Te$_{3}$, and Sb$_{2}$Te$_{3}$, displaying an anisotropic crystal structure composed by stacked quintuple layers (five covalently-bonded atomic layers) stabilized by van der Waals~\cite{Geim_13} interactions. With the aid of nanofabrication techniques such as vapor-phase growth~\cite{Peng_10,Kong_10,Kong_10B,Gehring_12}, solution-phase growth~\cite{Kong_13,Xiu_11,Purkayastha_06}, mechanical~\cite{Checkelsky_11,Kim_12,Seung_10} and chemical~\cite{Nicolosi_13} exfoliation, these materials allow for the synthesis of TI nanostructures at different sizes and morphologies~\cite{Parra_17,Goncalves_18,Fang_12,Chou_18,Kong_10B,Kong_13,Gehring_12}, including nanoribbons~\cite{Peng_10,Seung_10,Fang_12}, nanowires~\cite{Kong_10}, nanorods~\cite{Purkayastha_06,Goncalves_18} and nanoplates~\cite{Kong_10B,Kong_13,Gehring_12,Parra_17,Chou_18} of variable thickness with sharp edges and corners.

The band structure of these materials has been accurately predicted by first-principles
calculations~\cite{Zhang_09}, in agreement with several ARPES studies~\cite{Chiatti_16} that confirm the existence of well defined topologically non-trivial surface states with Dirac-like dispersion. In addition, magnetotransport experiments~\cite{Tang_19,Liao_17,Tian_13,Steinberg_11,Zhao_13,Kim_12,Chiatti_16,Checkelsky_11} seem to indicate that in many TI nanostructures the charge carriers correspond to both surface and bulk electronic states~\cite{Liao_17,Chiatti_16,Steinberg_11,Zhao_13,Checkelsky_11}, and that topologically protected surface states coexist and even interact with trivial surface states. The detailed mechanisms involving the coupling of bulk and trivial surface states to topologically protected surface states in such nanostructures are not fully understood yet, but are believed to originate on a combination of geometric confinement and disorder localization effects~\cite{Liao_17,Chiatti_16,Tian_13,Lu_11,Kim_12,Checkelsky_11}. In this article, we shall explore the origin of topologically trivial surface states as a consequence of morphological-dependent confinement in TI nanostructures, in order to shed some further light into such mechanisms. 

A combination of band structure calculations in TIs with a full group-theoretical analysis of their symmetries ~\cite{Liu_10,Zhang_09} leads to analytical effective low-energy models based on the $\mathbf{k}\cdot\mathbf{p}$ expansion near the $\Gamma$ point~\cite{Liu_10}. These minimal continuum low-energy models have been used to study the confinement effects over the single-particle spectrum in 3D TI nanoparticles~\cite{Imura_12,Governale_20} (TI-NP). Idealized geometries, such as spheres
~\cite{Imura_12} and infinitely long cylindrical rods ~\cite{Governale_20} have been studied so-far in the literature, and selection rules for optical transitions have been determined from analytical solutions in such cases. In these studies, geometrical confinement is mathematically enforced by a hard-wall condition~\cite{Imura_12,Governale_20}, thus demanding for the spinor eigenstates to vanish at the boundary of the domain representing the surface of the TI-NP, and hence by construction topologically trivial surface states are automatically ruled out in those analytical models. It is important to notice that, despite this is a valid boundary condition for the Schr\"odinger equation~\cite{castano2019impact,CASTANOYEPES2020114202}, it is not the more general one for the Dirac equation. In more technical mathematical terms, this problem corresponds to the search for an appropriate self-adjoint extension of the Hamiltonian operator~\cite{thaller2013dirac} within the domain imposed by the TI-NP geometry. Indeed, in physical terms the most general confinement condition involves the vanishing of the normal component of the probability current at the surface, as was long ago discussed in the context of the M.I.T bag model~~\cite{Johnson_75,Chodos_74A,Chodos_74B,Chodos_74C}, and this does not necessarily imply for the whole spinor eigenstate to vanish as well, as assumed in the hard-wall boundary condition applied in previous models~\cite{Imura_12,Governale_20}. Since the effective low-energy Hamiltonian that better describes TI-NPs~\cite{Zhang_09,Liu_10} involves a combination of linear (Dirac-like) and quadratic (Schr\"odinger-like) in momentum dispersions, it is important to obtain the corresponding mathematical expression for the probability current, in order to identify the appropriate boundary condition that determines confinement in the more general sense, given the morphological features of the TI-NP boundary.

In this work, we shall focus on TI-NPs with triangular shapes, as depicted in Fig.~\ref{fig:cuña}, as an approximation to some of the experimental morphologies observed in TI nanoplates involving sharp wedge angles~\cite{Parra_17,Goncalves_18,Chou_18,Kong_10B,Kong_13,Gehring_12}. For this particular geometry, we define the thickness by $h$, the radius by $R$ and the wedge angle by $\alpha$. Our goal is to investigate the energy spectrum of the confined bulk states within such TI-NPs,
as well as the corresponding probability density distribution, in order to identify the morphological features leading to the presence of topologically trivial surface states. We shall also study the corresponding selection rules for optical transitions between such confined states. For this purpose, as we further show, we developed an exact analytical solution to the eigenvalue problem, that allows us to investigate the scaling of the spectrum, density distribution and transition probabilities as a function of the morphological confinement parameters. Finally, using our explicit analytical solutions, we explore the features of a TI-NP with the shape of a very long cylindrical rod~\cite{Purkayastha_06,Goncalves_18} in the limit $\alpha\rightarrow 2\pi$, $h\rightarrow\infty$.

\section{The bulk Hamiltonian}
\begin{figure}
    \centering
    \includegraphics[scale=0.5]{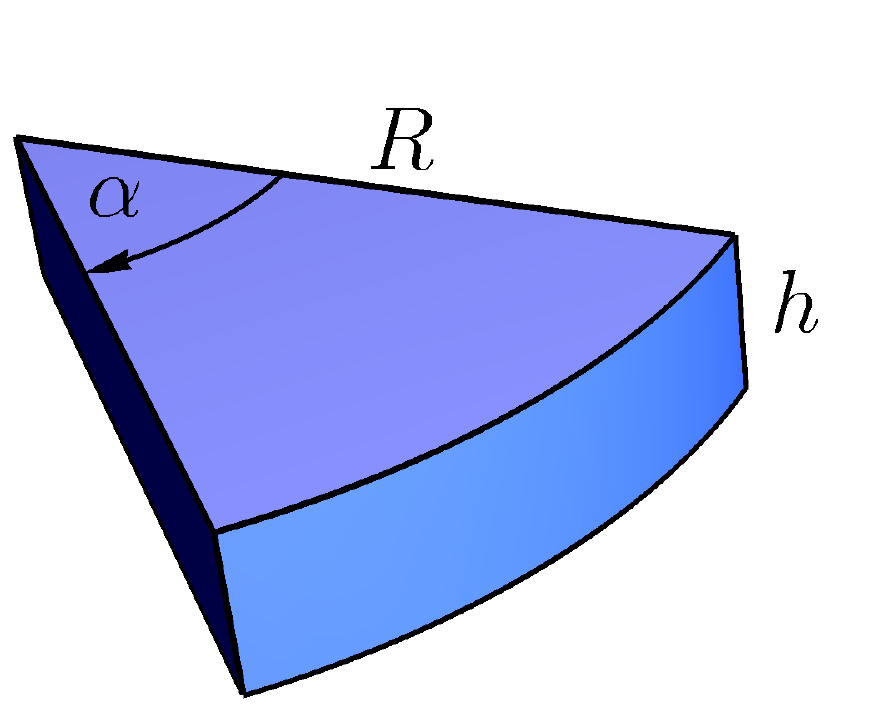}
    \caption{Sketch of the triangular-shaped TI nanoparticle of radius $R$ and height $h$, with a wedge angle $\alpha$.}
    \label{fig:cuña}
\end{figure}

We are interested in the study of confined electronic states within a TI-NP, whose geometry is depicted in Fig.~\ref{fig:cuña}, and hence we start from the corresponding bulk Hamiltonian for the material. For this purpose, we consider a low-energy continuum model that represents the band structure of a typical topological insulator near the $\Gamma$-point~~\cite{Liu_10,Zhang_09,Imura_12}:
\bea
\Ham&=&\left(\frac{\Delta_0}{2}+\frac{\gamma}{k_\Delta}\mathbf{k}^2\right)\tau_z\otimes\sigma_0\nn\\
&+&\gamma\left(k\tau_x\otimes\sigma_z+k_-\tau_x\otimes\sigma_++k_+\tau_x\otimes\sigma_-\right),
\label{Hamiltonian}
\eea
where we have taken into account the spin ($\sigma$) and pseudo-spin ($\tau$) degrees of freedom. Here, we defined the momentum operators $\mathbf{k}=(k_x,k_y,k)=-\ii\nabla$, $k_\pm=k_x\pm \ii k_y$, as well as the ladder spin operators $\sigma_\pm=(\sigma_x\pm\ii\sigma_y)/2$. The numerical values for the parameters $\Delta_0$, $\gamma$, and $k_{\Delta}$ are presented in Table~\ref{tab:parameters} for three different TI materials.

Based on this Hamiltonian model for the bulk, we incorporate the confinement effects induced by the specific TI-NP geometry by imposing a current insulating condition at the surface~~\cite{Johnson_75,Chodos_74A,Chodos_74B,Chodos_74C}. For a confined state,
only the {\it{normal}} component of the current $\left.\mathbf{n}\cdot\mathbf{J}\right|_{\partial\Omega}$ determines the net probability flux through the boundary $\partial\Omega$, and therefore it must vanish at each of the surfaces depicted in Fig.~\ref{fig:cuña}, i.e. at $\phi=\{0,\alpha\}$, $z=\{0,h\}$, and $r=R$. The later yields quantization conditions, as we shall discuss in further detail, and the correct identification of the probability density at the surface of the system. 
In order to simplify the mathematical notation, and further algebraic manipulations, we define the "mass" differential operator
\bea
\mathbb{M}(\mathbf{k})\equiv\frac{\Delta_0}{2}+\frac{\gamma}{k_\Delta}\mathbf{k}^2,
\eea
so that the Hamiltonian Eq.~\eqref{Hamiltonian} can be written in the alternative and more compact form
\bea
\Ham=\mathbb{M}(\mathbf{k})\hat{\beta}+\gamma\left(\boldsymbol{\alpha}\cdot\mathbf{k}\right),
\label{eq:Ham2}
\eea
where we defined the Dirac $4\times4$ matrices in the standard representation
\bea
\boldsymbol{\alpha}=\begin{pmatrix}
0 & \boldsymbol{\sigma}\\
\boldsymbol{\sigma} & 0\end{pmatrix},~~~\hat{\beta}=\begin{pmatrix}
\mathbb{I} & 0\\
0& -\mathbb{I}\end{pmatrix}.
\eea
Therefore, with the purpose of studying the energy spectrum and confined states within the nanoparticle, we investigate the eigenvalue problem $\Ham \Psi (r,\phi,z) = E \Psi(r,\phi,z)$, where $\Psi = \left(\varphi,\chi\right)^\text{T}$ is a 4-component spinor. This eigenvalue problem, according to the matrix structure in Eq.~\eqref{eq:Ham2}, leads to the coupled system of differential equations
\begin{subequations}
\bea
\left[\mathbb{M}(\mathbf{k})-E\right]\varphi+\gamma\left(\boldsymbol{\sigma}\cdot\mathbf{k}\right)\chi=0,
\label{eq1}
\eea
\bea
\gamma\left(\boldsymbol{\sigma}\cdot\mathbf{k}\right)\varphi-\left[\mathbb{M}(\mathbf{k})+E\right]\chi=0.
\label{eq2}
\eea
\end{subequations}

Based on the explicit analytical solutions obtained for the constant mass limit $\mathbb{M}(\mathbf{k}^2)\rightarrow \Delta_0/2$ (see Sec. 2 in Supplemental material for details), we solve the general case by the spinors
\begin{subequations}
\bea
\varphi=e^{\ii kz}\begin{pmatrix}
     c_1 e^{\ii j\phi}J_j(\kappa r)\\
     \\
     c_2e^{\ii(j+1)\phi}J_{j+1}(\kappa r)
\end{pmatrix},
\eea
\bea
\chi=e^{\ii kz}\begin{pmatrix}
     c_3 e^{\ii j\phi}J_j(\kappa r)\\
     \\
     c_4e^{\ii(j+1)\phi}J_{j+1}(\kappa r)
\end{pmatrix},
\eea
\label{eq:ansatz}
\end{subequations}
where $J_{j}(x)$ are Bessel functions of order $j$, and $\kappa$ is a real constant, to be determined as a function of the energy $E$, as follows.

\begin{table*}
\centering
\begin{tabular}{ |p{3cm}|p{2cm}|p{2cm}|p{2cm}|p{2cm}|  }
 \hline
 Material& $\Delta_0$~(eV)  & $\gamma$~(eV \r{A}) & $k_{\Delta}$~($\text{\r{A}}^{-1}$) & $R_0$~(\r{A}) \\
 \hline
 Bi$_2$ Se$_{3}$~~\cite{Nechaev_16}   & -0.338  &2.2&   0.13 & 1.3\\
 Bi$_2$ Te$_{3}$~~\cite{Liu_10} &   -0.3  & 2.4   &0.026 & 0.81 \\
 Pb$_{0.81}$ Sn$_{0.19}$ Se~~\cite{Assaf_17}  &-0.025 & 3.2&  0.2& 26\\
 \hline
\end{tabular}
\caption{\label{tab:parameters}Parameters for the bulk Hamiltonian Eq.~\eqref{Hamiltonian} for different topological materials, after Refs.~~\cite{Liu_10,Nechaev_16,Assaf_17}.}
\end{table*}

By inserting the spinors Eq.~\eqref{eq:ansatz} into Eq.~\eqref{eq1} and Eq.~\eqref{eq2}, we obtain the algebraic linear system
\bea
\mathbb{A}(\kappa,E)\mathbf{c}=0,
\label{systemforc}
\eea
where we defined the vector $\mathbf{c}=\left(c_1,c_2,c_3,c_4\right)^\text{T}$, as well as the matrix
\bea
\mathbb{A}(\kappa,E)=\begin{bmatrix}
\Lambda-E & 0 & k\gamma &-\ii\kappa\gamma\\
0 & \Lambda-E & \ii\kappa\gamma &-k\gamma\\
k\gamma & -\ii\kappa\gamma & -(\Lambda+E) & 0\\
\ii\kappa\gamma & -k\gamma & 0 &-(\Lambda+E)
\end{bmatrix}.
\eea

Here, we also defined the coefficient
\bea
\Lambda=\frac{\Delta_0}{2}+\frac{\gamma}{k_\Delta}\left(k^2+\kappa^2\right).
\eea

The linear system is solvable if $\det\,\mathbb{A}=0$, which implies the secular equation
\bea
\left[\Lambda^2-E^2+\gamma^2(k^2+\kappa^2)\right]^2=0,
\eea

whose solutions are the energy eigenvalues as a function of $\kappa$ 
\bea
E_\pm=\pm\sqrt{\left(\frac{\Delta_0}{2}+\frac{\gamma}{k_\Delta}\left(k^2+\kappa^2\right)\right)^2+\gamma^2(k^2+\kappa^2)}.
\label{eq:Ekappa}
\eea
\subsection{Spinor eigenvectors and energy spectrum}
The linear system defined by Eq.~(\ref{systemforc}) possesses four independent eigenvectors defined by the coefficients $\mathbf{c} = \left(c_1,c_2,c_3,c_4  \right)^T$, as follows
\bea
\begin{pmatrix}
     \ii\frac{\kappa\left(-\Lambda\pm\sqrt{\Lambda^2+\gamma^2(k^2+\kappa^2)}\right)}{\gamma(k^2+\kappa^2)}\\
     \\
     \frac{k\left(-\Lambda\pm\sqrt{\Lambda^2+\gamma^2(k^2+\kappa^2)}\right)}{\gamma(k^2+\kappa^2)}\\
     \\
     0\\
     \\
     1
\end{pmatrix},
\begin{pmatrix}
     \frac{k\left(\Lambda\pm\sqrt{\Lambda^2+\gamma^2(k^2+\kappa^2)}\right)}{\gamma(k^2+\kappa^2)}     \\
     \\
\ii\frac{\kappa\left(\Lambda\pm\sqrt{\Lambda^2+\gamma^2(k^2+\kappa^2)}\right)}{\gamma(k^2+\kappa^2)}\\
     \\
     1\\
     \\
     0
\end{pmatrix}.\nn\\
\label{eq:cvec}
\eea

By construction, these four solutions are clearly degenerated by pairs, as seen in Eq.~\eqref{eq:Ekappa}. However, we still must determine the explicit values for the parameters $\kappa$, $k$, and $j$, and here is where the geometrical confinement effects come into place. We show that the probability current associated to the Hamiltonian Eq.~\eqref{Hamiltonian} is given by the expression (see Sec. 1 in Supplemental for details)
\bea
\mathbf{J}&=&\frac{\ii\gamma}{k_\Delta}\left[\nabla\Psi^\dagger\left(\tau_z\otimes\sigma_0\right)\Psi-\Psi^\dagger\left(\tau_z\otimes\sigma_0\right)\nabla\Psi\right]\nn\\
&+&\Psi^\dagger\boldsymbol{\alpha}\Psi
\label{J},
\eea
where the first contribution arises from the quadratic dispersion terms in the Hamiltonian, and hence is reminiscent from the standard Schr\"odinger expression, while the second term arises from the linear terms, leading to a Dirac pseudo-relativistic expression.

As we already pointed out, the mathematical condition representing quantum confinement of the electronic eigenstates within the TI-NP is that the normal current at the boundary $\partial\Omega$ vanishes, i.e. $\left.\mathbf{n}\cdot\mathbf{J}\right|_{\partial\Omega} = 0$. By imposing this condition at the surfaces $z = 0$ and $z = h$, i.e. $\hat{\mathbf{e}}_z\cdot\left.\mathbf{J}\right|_{z=\{0,h\}}=0$, we are led to construct linear combinations from the basic solutions of the form $e^{i k z} - e^{-i k z} \propto \sin(k z)$,
to obtain a quantization of the $z$-component of the momentum $k$
\begin{equation}
k = \frac{m\pi}{h},\,m = 1, 2,\ldots.
\label{k_quant}
\end{equation}

The analogous condition at the lateral surfaces of the wedge $\phi = 0$ and $\phi = \alpha$, i.e. $\hat{\mathbf{e}}_{\phi}\cdot\left.\mathbf{J}\right|_{\phi=\{0,\alpha\}}=0$, leads to linear combinations of the basic solutions of the form $e^{i j \phi} - e^{-i j \phi} \propto \sin(j\phi)$, with the corresponding quantization of the index $j$
\begin{eqnarray}
j = \frac{l\pi}{\alpha},\,l = 1, 2, \ldots
\label{j_quant}
\end{eqnarray}

\begin{figure*}[t!]
    \centering
    \includegraphics[scale=0.5]{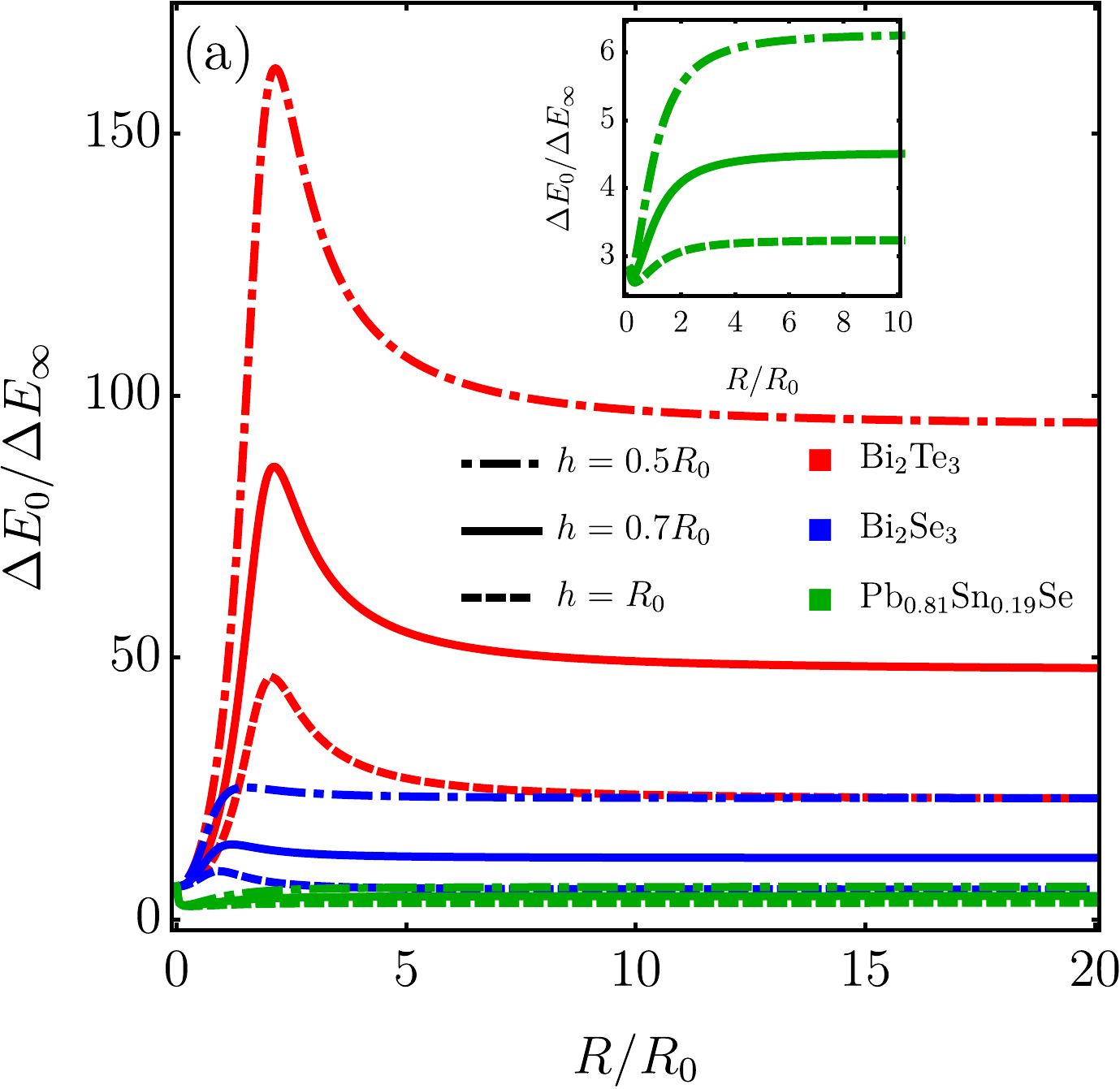}\hspace{0.5cm}\includegraphics[scale=0.5]{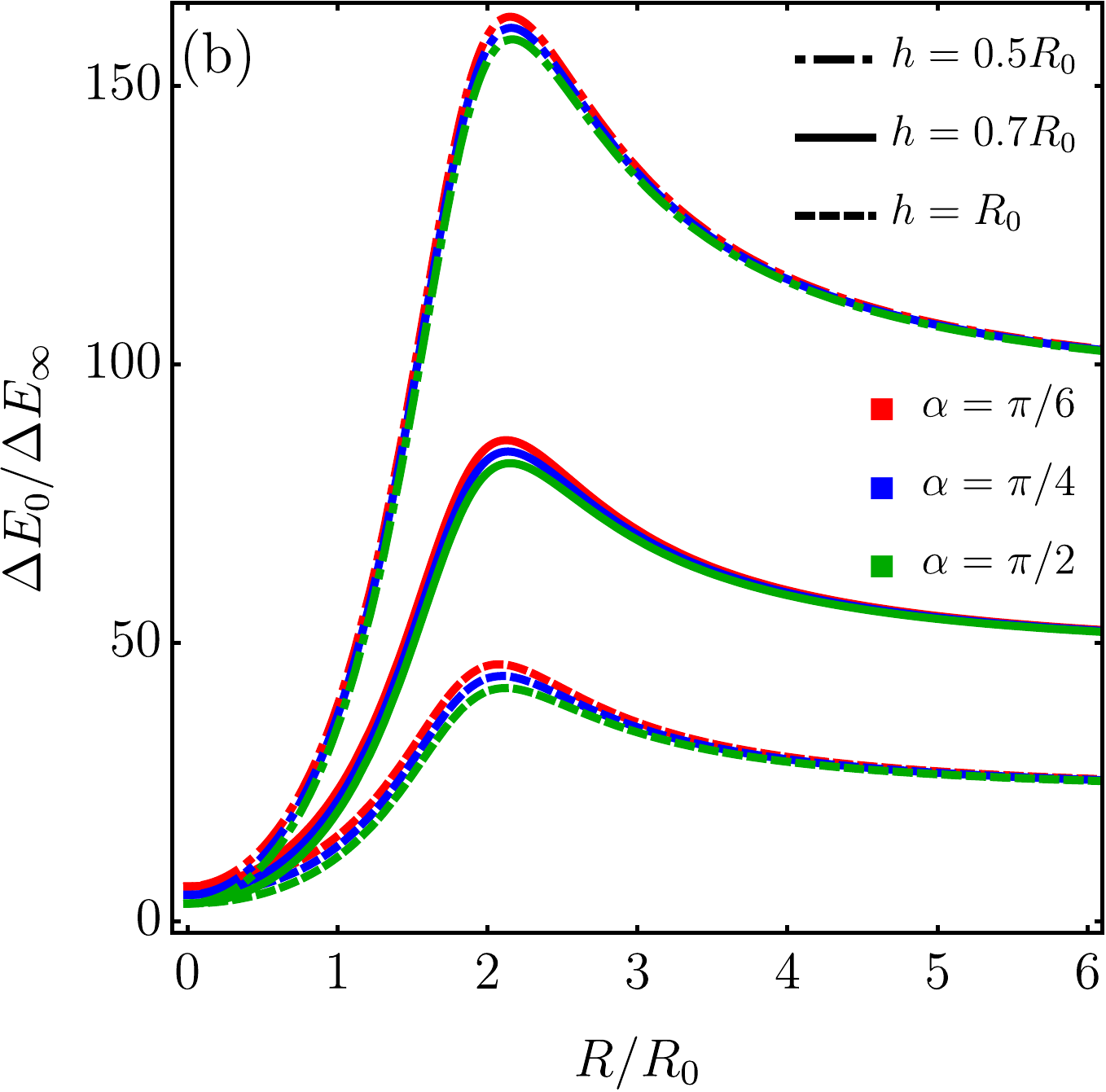}
    \caption{Energy gap for TI nanoparticles of different materials, as a function of the 
    morphological parameters $\alpha$, $h$, and $R$. (a) Bi$_2$Te$_3$, Bi$_2$Se$_3$, and Pb$_{0.81}$Sn$_{0.19}$Se, respectively, at fixed $\alpha = \pi/6$. The inset shows the impact of the material parameters on the energy scaling for Pb$_{0.81}$Sn$_{0.19}$Se. (b) Bi$_2$Te$_3$ for three different values of $\alpha = \pi/6$, $\pi/4$, and $\pi/2$, respectively. Here $R_0=2\gamma/|\Delta_0|$ and $\Delta E_{\infty}$ is given by Eq.~(\ref{DeltaER_2limits}).}
    \label{fig:DeltaE0_2limits}
\end{figure*}

By incorporating the quantization conditons Eq.~\eqref{k_quant} and Eq.~\eqref{j_quant}, in order to simplify the notation, let us define the coefficients
\bea
\mathcal{M}&=&\frac{\Lambda}{\gamma(k^2+\kappa^2)},\nn\\
\widetilde{\mathcal{M}}&=&\frac{\sqrt{\Lambda^2+\gamma^2(k^2+\kappa^2)}}{\gamma(k^2+\kappa^2)},
\eea
such that the four independent states, combining Eq.~\eqref{eq:cvec} and Eq.~\eqref{eq:ansatz}, are given by:
\begin{subequations}
\bea
\langle\mathbf{r}|nml;1^\pm\rangle&=&\sin(kz)\sin\left(j\phi\right)\nn\\
&\times&\begin{pmatrix}
     \ii\kappa(-\mathcal{M}\pm\widetilde{\mathcal{M}}) J_j(\kappa r)\\
     \\
     k(-\mathcal{M}\pm\widetilde{\mathcal{M}}) J_{j+1}(\kappa r)e^{\ii\phi}\\
     \\
     0\\
     \\
     J_{j+1}(\kappa r)e^{\ii\phi}
\end{pmatrix},
\eea
\bea
\langle\mathbf{r}|nml;2^\pm\rangle&=&\sin(kz)\sin\left(j\phi\right)\nn\\
&\times&\begin{pmatrix}
     k(\mathcal{M}\pm\widetilde{\mathcal{M}}) J_j(\kappa r)\\
     \\
     \ii\kappa(\mathcal{M}\pm\widetilde{\mathcal{M}}) J_{j+1}(\kappa r)e^{\ii\phi}\\
     \\
     J_j(\kappa r)\\
     \\
     0
\end{pmatrix},
\eea
\end{subequations}
and the general solution can be written as the linear combination
\bea
\ket{nml}=\sum_{s=\pm}\left[A_s\ket{nml,1^s}+B_s\ket{nml,2^s}\right].
\eea

Finally, imposing the confinement condition at the external radial surface $r = R$, i.e.
$\left.\hat{\mathbf{e}}_r\cdot\mathbf{J}\right|_{r=R} = 0$ leads to the
equation
\bea
J_{\frac{l\pi}{\alpha}}(\kappa R)J_{\frac{l\pi}{\alpha}+1}(\kappa R)=0,
\label{conditionforkappa}
\eea
where the quantization conditon Eq.~\eqref{j_quant} over $j$ was explicitly implemented in terms of the integer $l$.
\begin{figure*}[t!]
    \centering
    \includegraphics[scale=0.3]{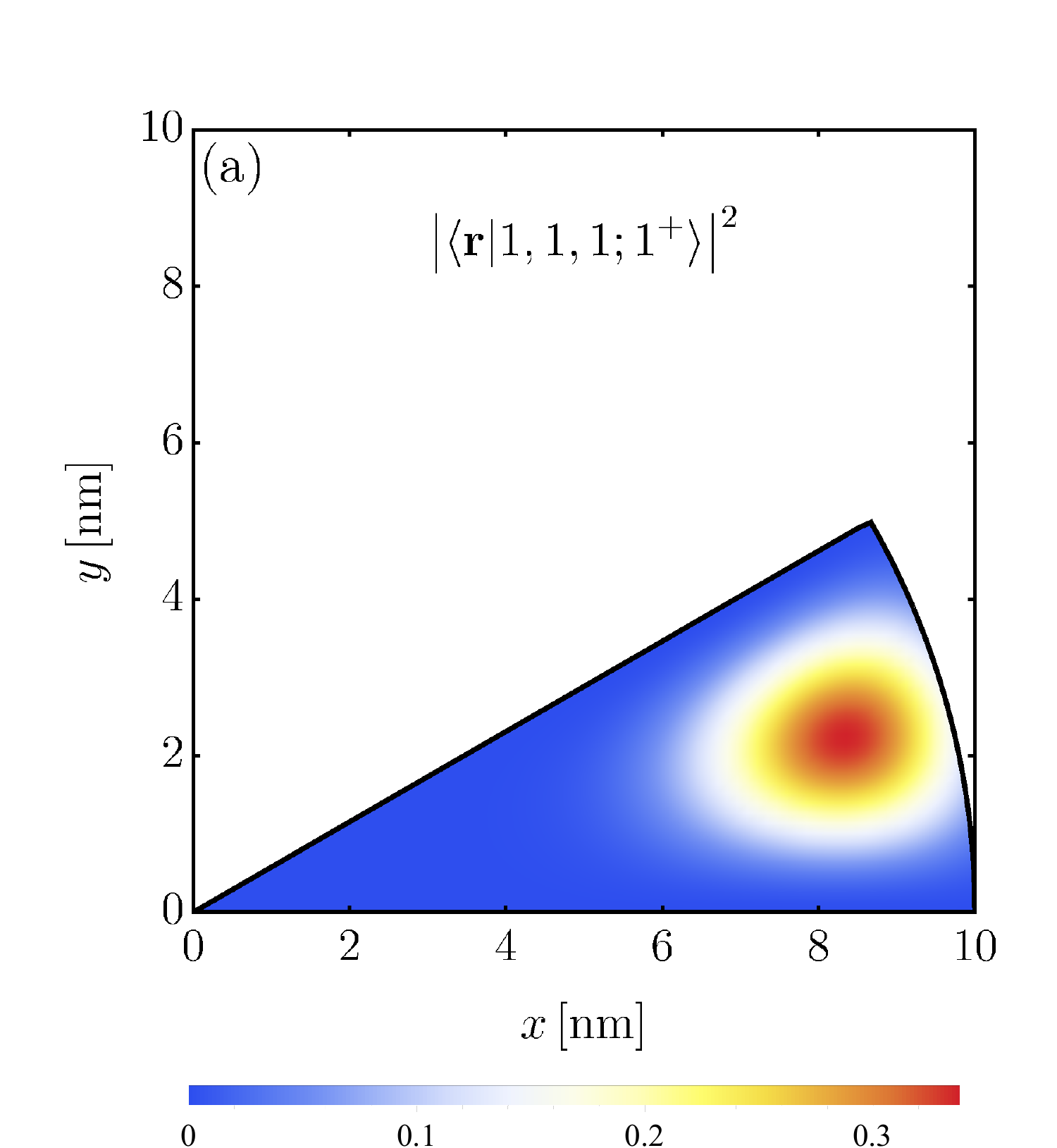}\includegraphics[scale=0.3]{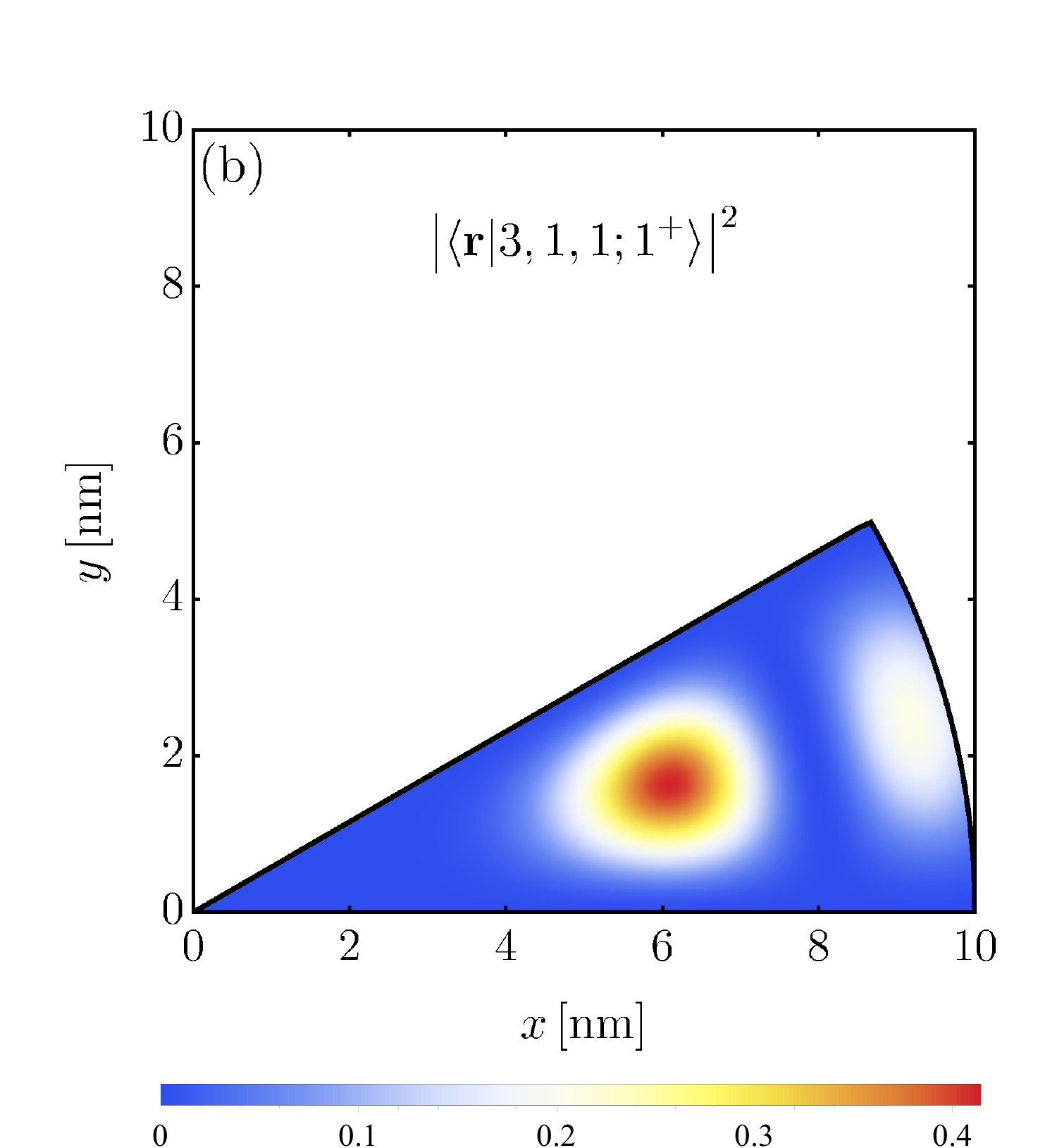}\includegraphics[scale=0.3]{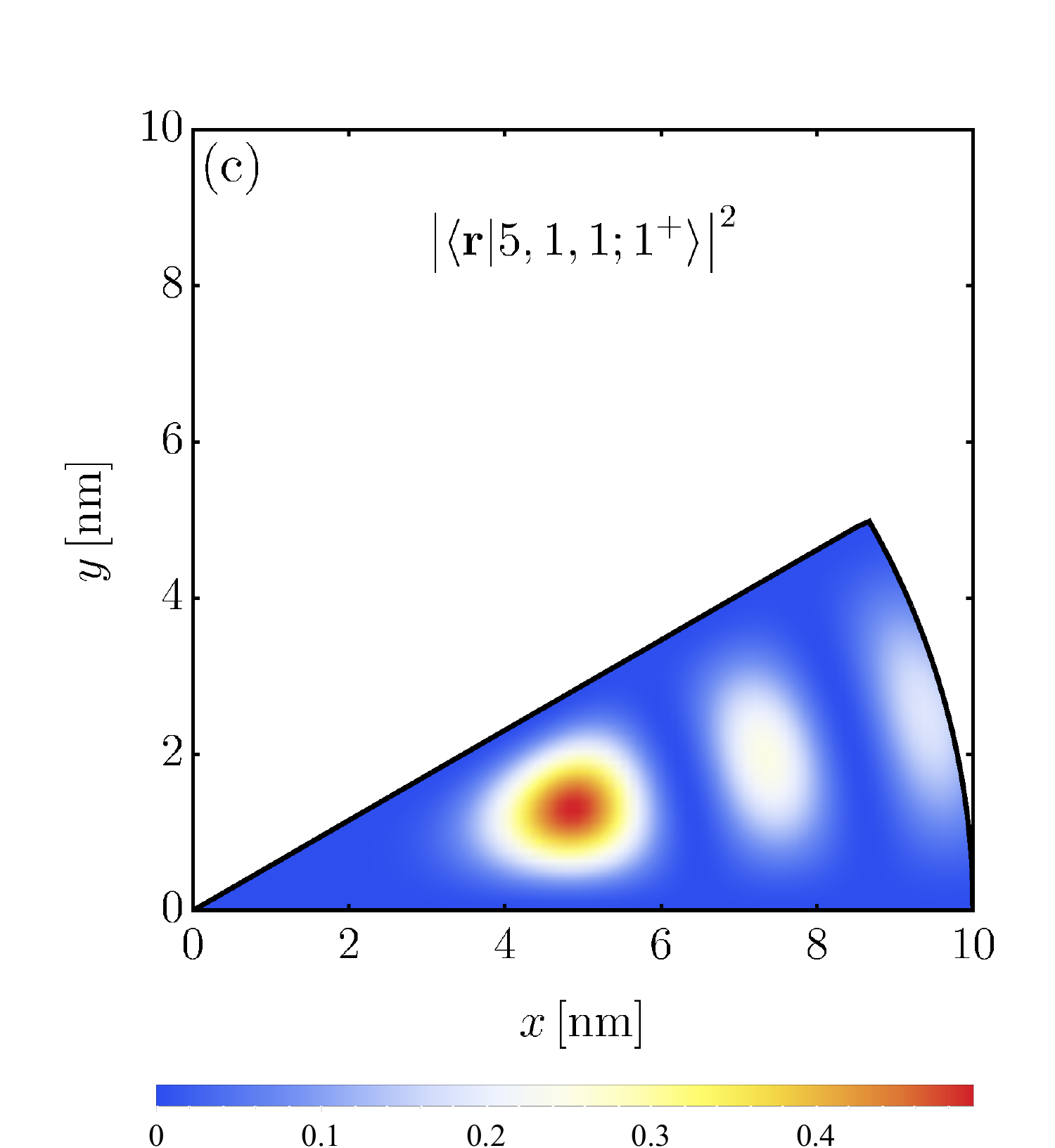}\\
    \includegraphics[scale=0.3]{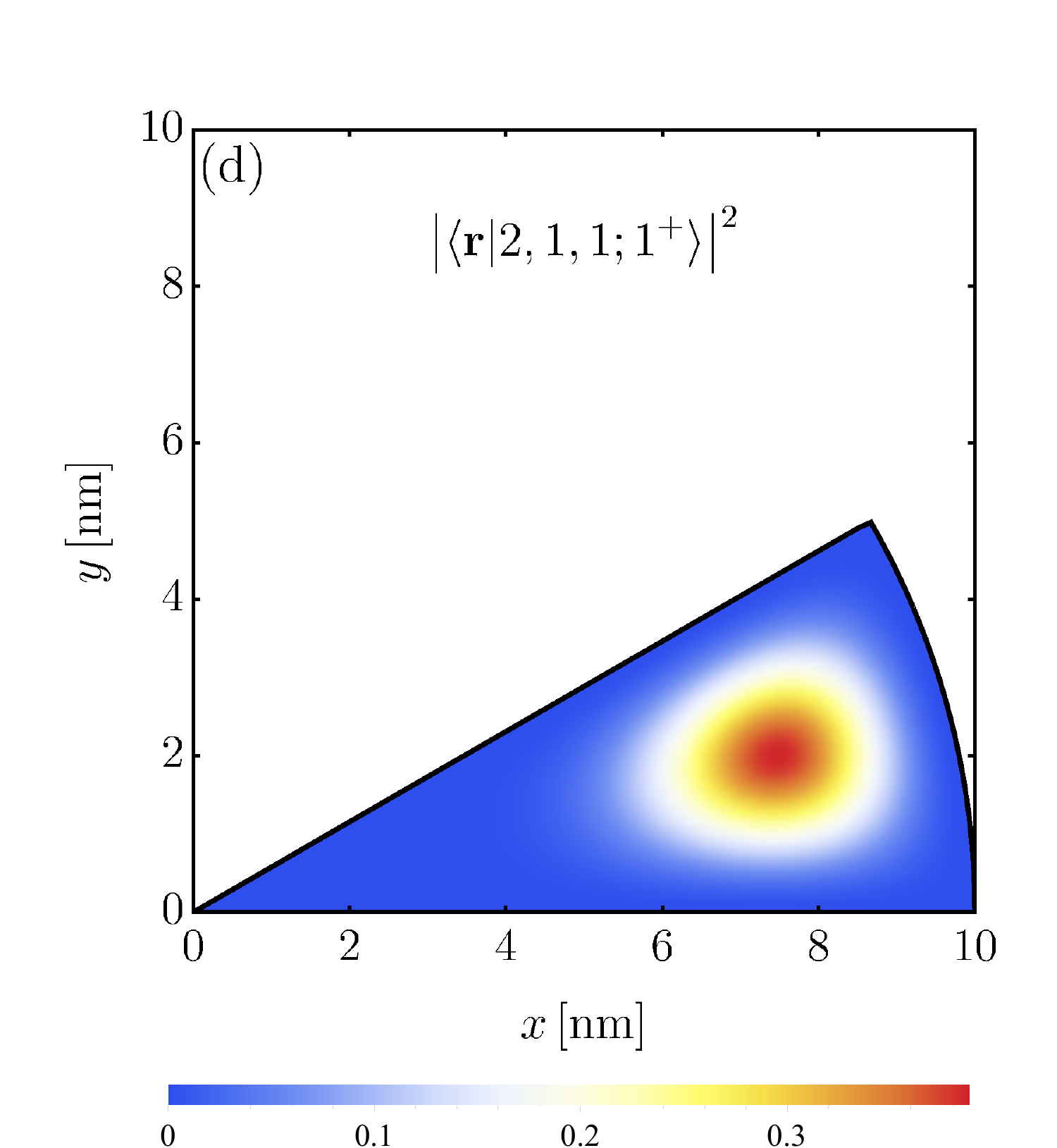}\includegraphics[scale=0.3]{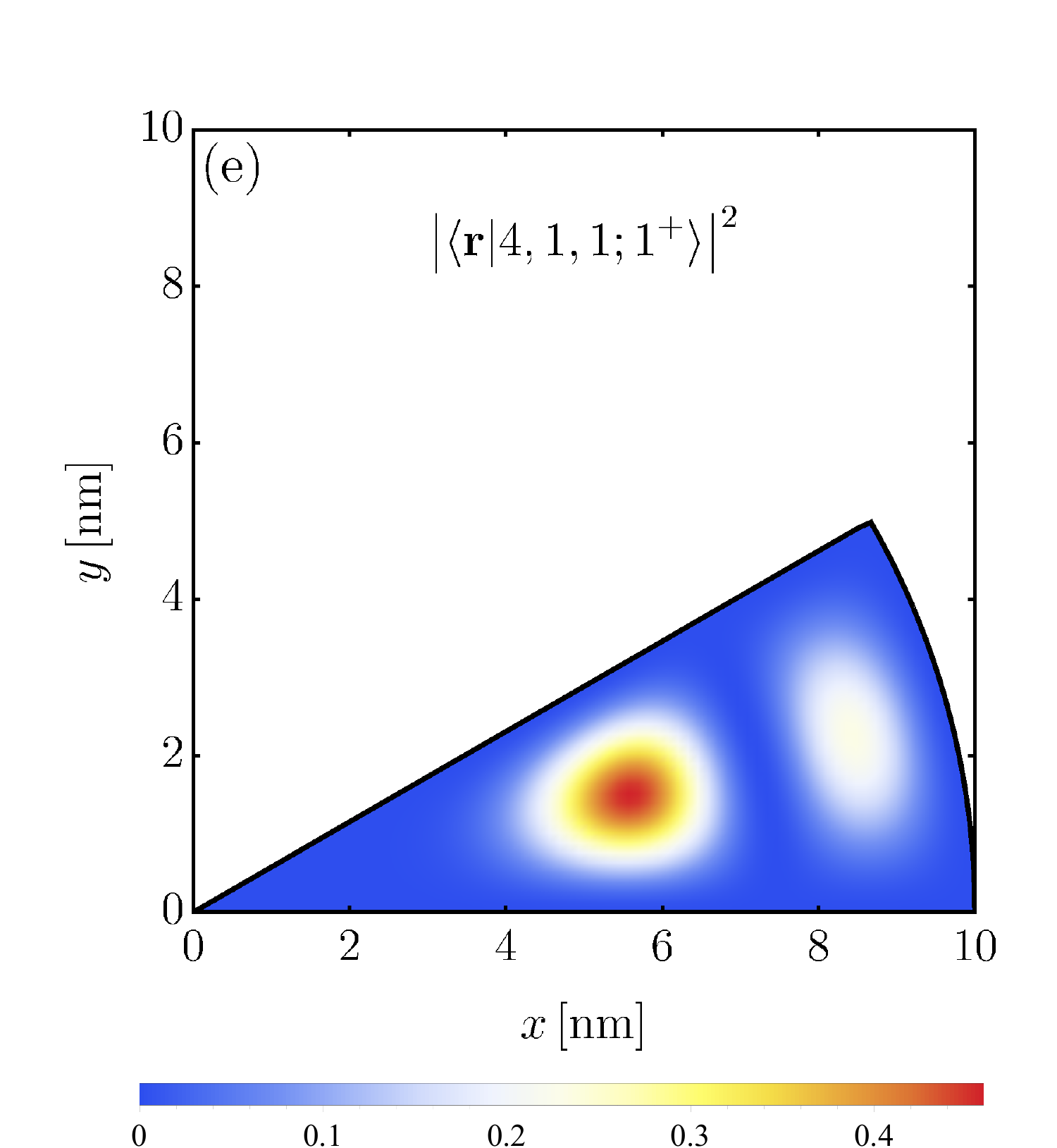}\includegraphics[scale=0.3]{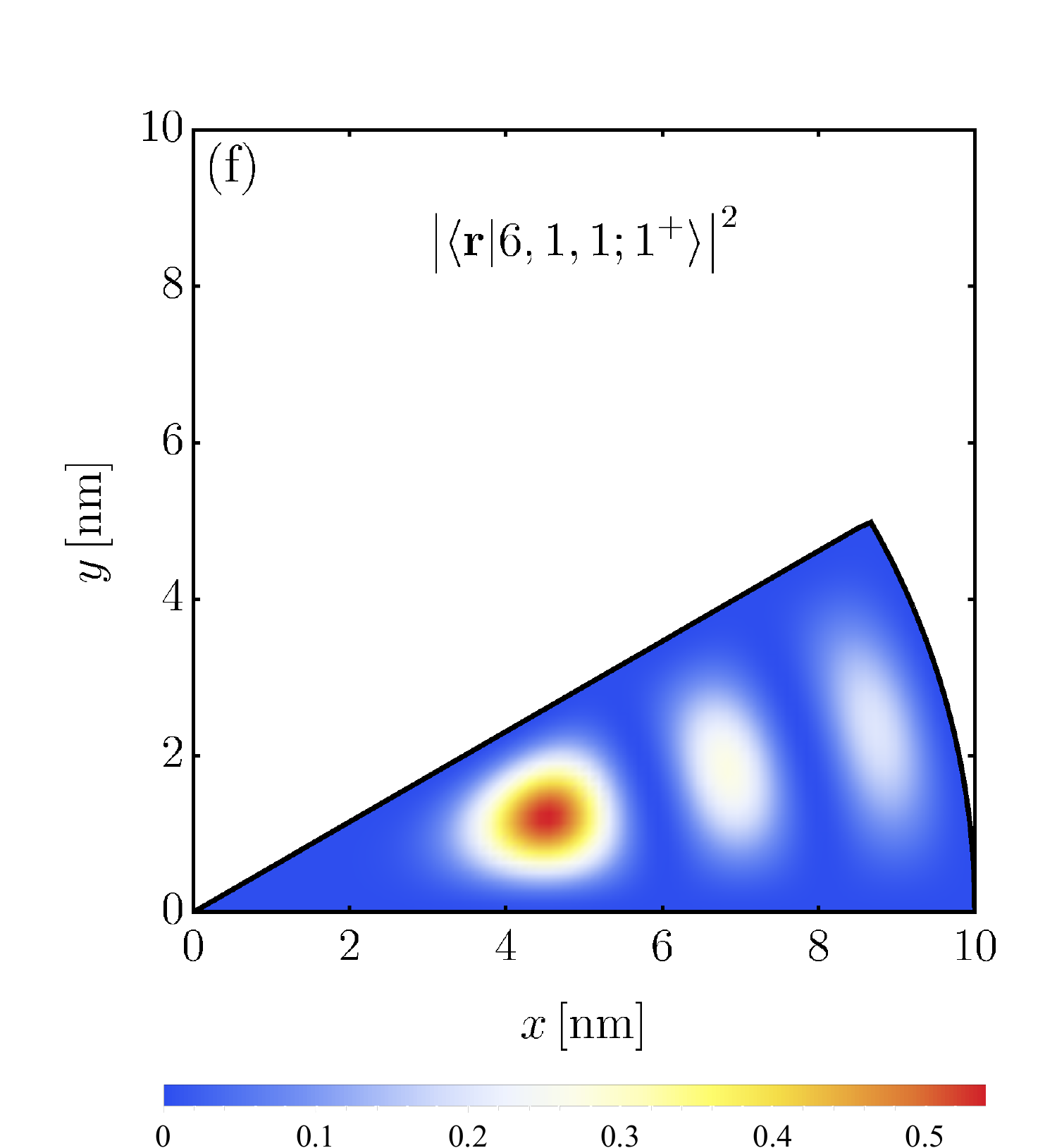}
    \caption{Probability density, represented as a 2D-color plot at the plane $z = 0.5 h$, for different confined states within a TI nanoparticle with $\alpha = \pi/6$, $l = m = 1$, and $1\le n \le 6$.}
    \label{fig:role_of_n_1up}
\end{figure*}

Therefore, if $\kappa_{nl} = \kappa R$ are the (infinitely many) roots of Eq.~(\ref{conditionforkappa}), the energy eigenvalues for the confined electronic states are given by the explicit analytical expression
\bea
E_{nml}^\pm&=&\pm\Bigg[\left[\frac{\Delta_0}{2}+\frac{\gamma}{k_\Delta}\left(\frac{m^2\pi^2}{h^2}+\frac{\kappa_{nl}^2}{R^2}\right)\right]^2\nn\\
&+&\gamma^2\left(\frac{m^2\pi^2}{h^2}+\frac{\kappa_{nl}^2}{R^2}\right)\Bigg]^{1/2}.
\eea

It is interesting to remark that this expression, in the limit of $h\rightarrow \infty$, restores the continuum limit for the $z$-component of the linear momentum (see also Eq.~\eqref{k_quant}). Moreover, as $\alpha\rightarrow 2\pi$ the wedge becomes a full cylinder, and as seen in Eq.~\eqref{j_quant} the usual total angular momentum quantization is restored, as $j$ is either an integer or half-integer in this limit, i.e. $j \rightarrow l-1/2$ for $l=0,\pm1,\pm2,\ldots$, and $\sin(j\phi) \rightarrow e^{\pm\ii(l-1/2)\phi}$. Therefore, in the limit of a full ($\alpha\rightarrow 2\pi$) infinitely long ($h\rightarrow\infty$) cylinder, the quantization condition Eq.~\eqref{conditionforkappa} reduces to
\bea
J_{l-1/2}(\tilde{\kappa} R)J_{l+1/2}(\tilde{\kappa} R)=0.
\eea

Let us define the energy gap for the infinitely long full cylinder by the expression
\bea
 \Delta E_{\infty}&=&\lim \limits_{\substack{%
    h \to \infty\\
    \alpha \to 2\pi}}\left( E_{1,0,0}^{+} - E_{1,0,0}^{-}\right)\nn\\
    &=& 2\sqrt{\left[\frac{\Delta_0}{2}+\frac{\gamma}{k_\Delta}\left(\frac{\tilde{\kappa}_{10}}{R}\right)^2\right]^2+\gamma^2\left(\frac{\tilde{\kappa}_{10}}{R}\right)^2}.
    \label{DeltaER_2limits}
\eea

In Fig.~\ref{fig:DeltaE0_2limits}, we represent the energy gap $\Delta E_0 = E^{+}_{1,1,1} - E^{-}_{1,1,1}$, defined as the difference between the lowest unoccupied and highest occupied energy levels, normalized by the corresponding $\Delta E_{\infty}$, as a function of the particle radius, in units of the "Compton wavelength" $R_0 = 2\gamma/|\Delta_0|$. We compare the energy gap for three different TI materials, Bi$_2$Te$_3$, Bi$_2$Se$_3$, and Pb$_{0.81}$Sn$_{0.19}$Se, respectively, as a function of the TI-NP height $h$ (in units of $R_0$) (Fig.~\ref{fig:DeltaE0_2limits}(a)), and for different values of the wedge angle $\alpha$ (Fig.~\ref{fig:DeltaE0_2limits}(b)). From Fig.~\ref{fig:DeltaE0_2limits}(a), it is evident that the TI-NP gap is extremely sensitive to the specific material parameters, specially the size of the bulk gap $\Delta_0$, which is comparatively larger for Bi$_2$Te$_3$. It is also clear that the confinement in the $z$-direction, represented by the TI-NP thickness $h$ is the most relevant at determining the size of the gap, as seen in both Fig.~\ref{fig:DeltaE0_2limits}(a,b). In particular, Fig.~\ref{fig:DeltaE0_2limits}(b) shows that the gap is very weakly dependent on the wedge angle $\alpha$. In all three examples, the changes in the TI morphology increases the energy gap (as compared with the infinite long cylndrical rod) up to two orders of magnitude for the material constants considered in Table~\ref{tab:parameters}.

\section{The probability density and topologically trivial surface states}

\begin{figure*}[t!]
    \centering
    \includegraphics[scale=0.3]{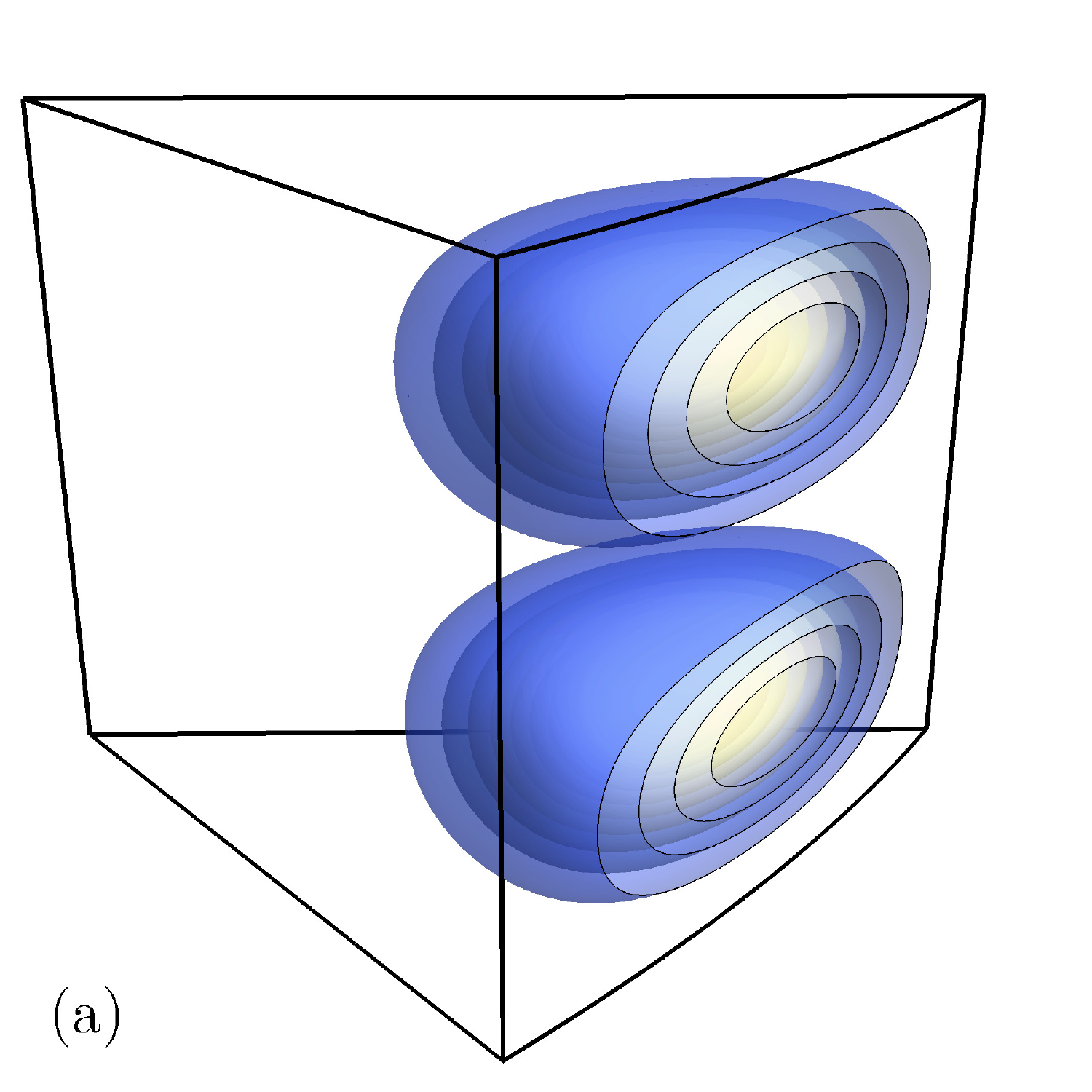}\hspace{0.5cm}\includegraphics[scale=0.3]{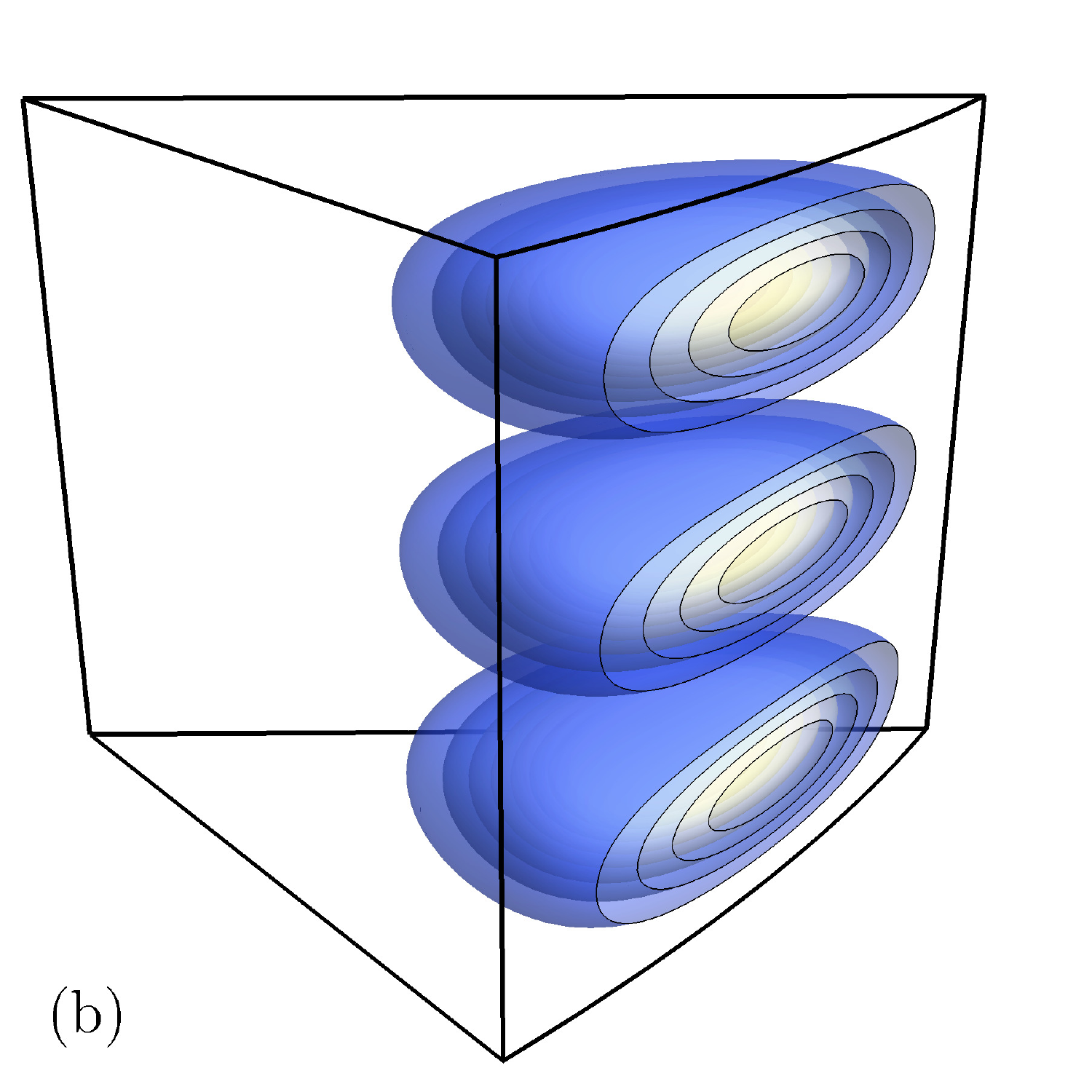}\\
    \includegraphics[scale=0.5]{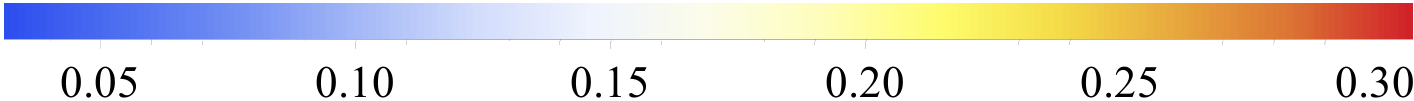}
    \caption{Probability density of the state $\ket{1,m,1;1^+}$ for (a) $m=2$ and (b) $m=3$ with $\alpha=\pi/6$.}
    \label{fig:role_of_m_1up}
\end{figure*}

\begin{figure*}[t!]
    \centering
    \includegraphics[scale=0.3]{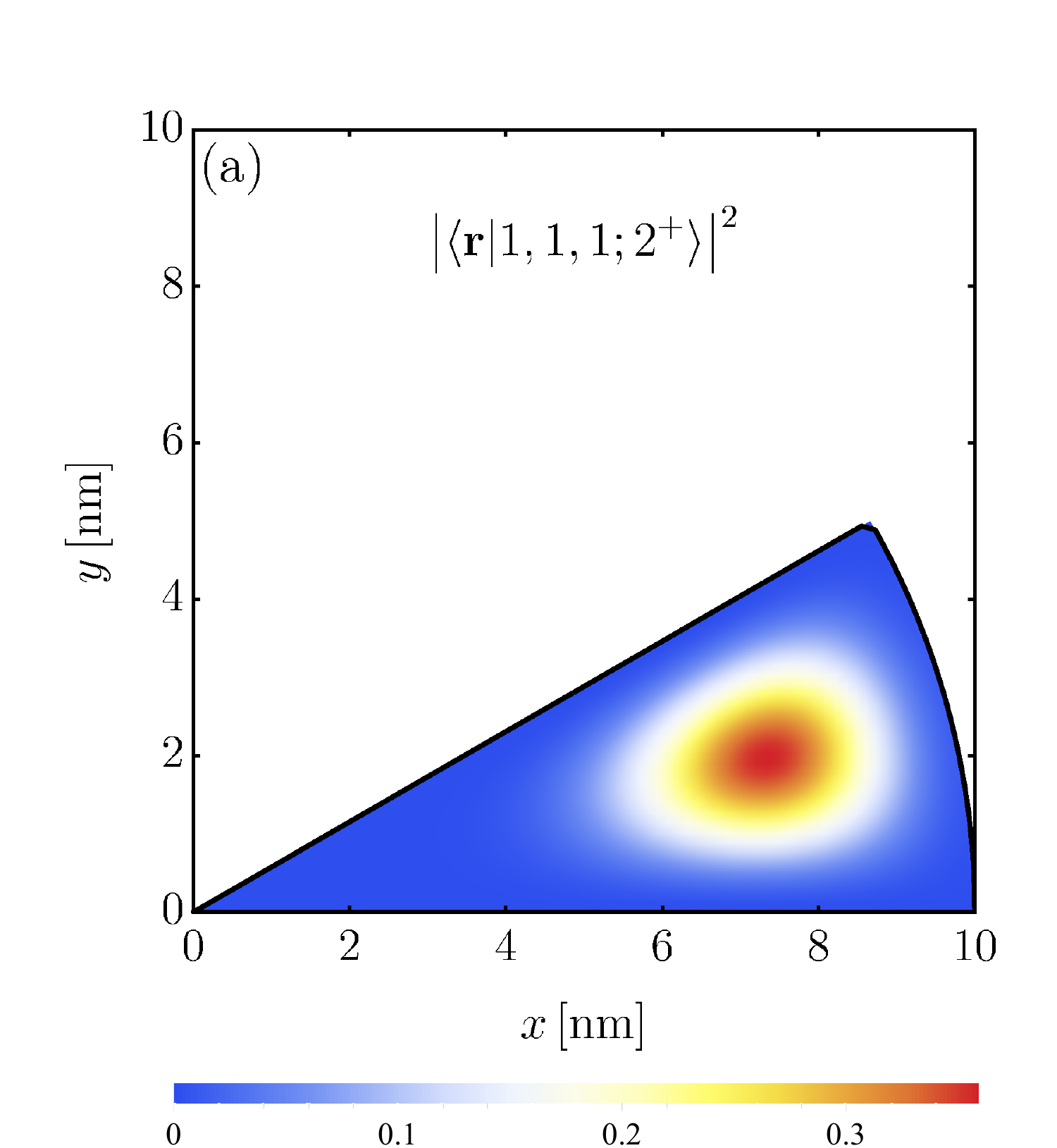}\hspace{0.7cm}\includegraphics[scale=0.3]{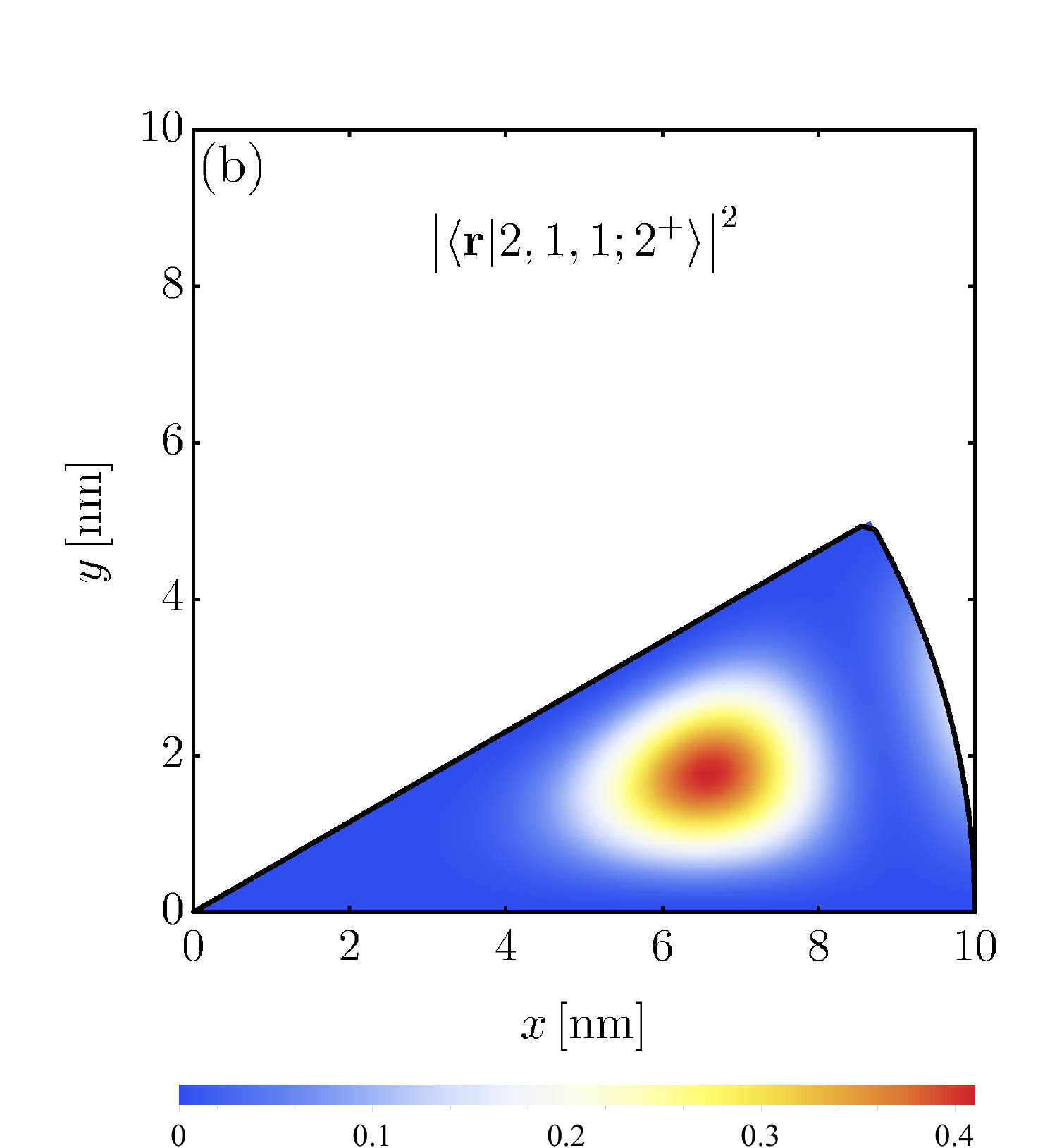}\hspace{0.5cm}\hspace{0.7cm}\includegraphics[scale=0.3]{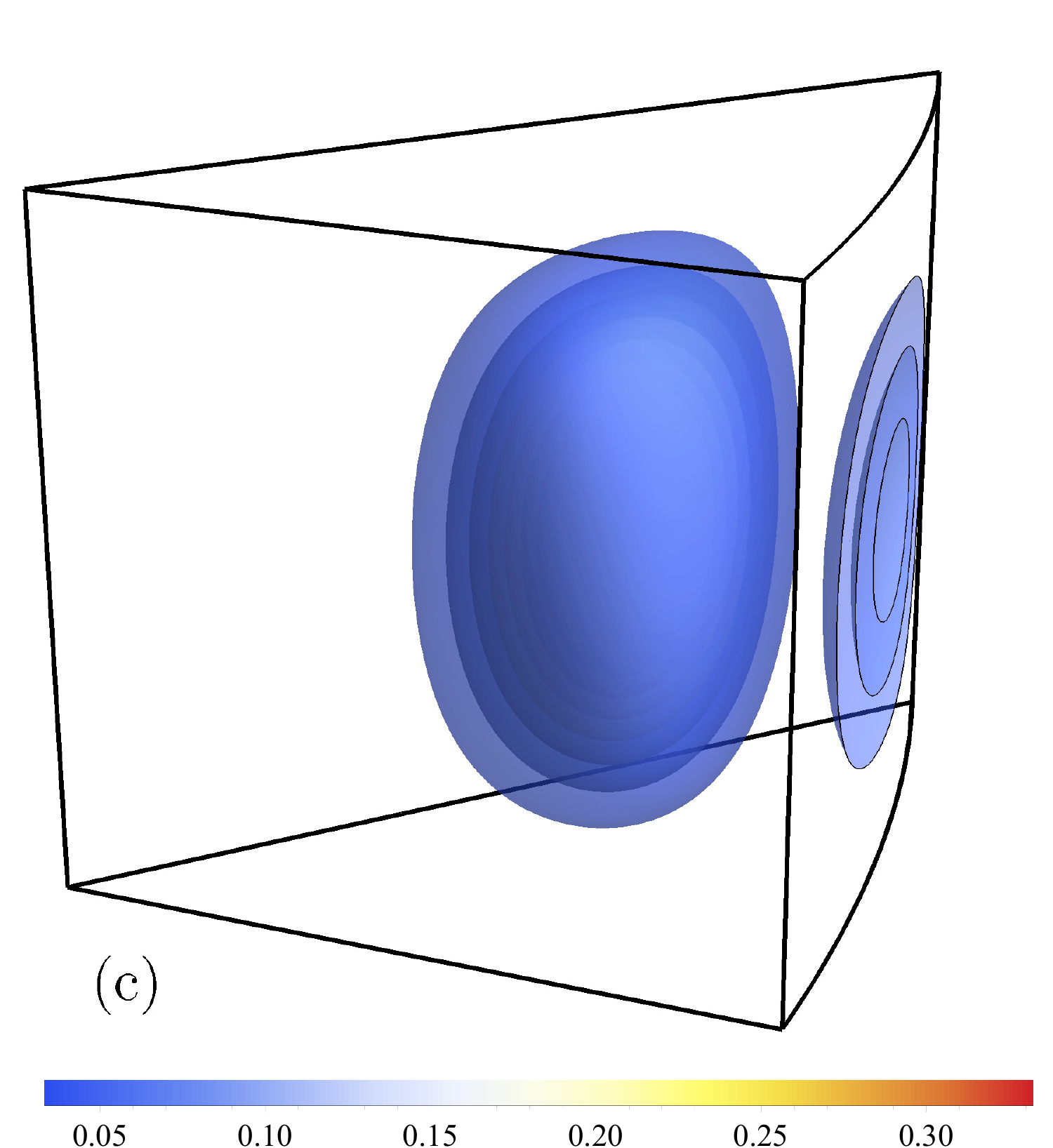}
    \caption{Probability density of the state $\ket{n,1,1;2^+}$ for (a) $n=1$ and (b) $n=2$ with $\alpha=\pi/6$ at $z=0.5h$. In order to appreciate the surface state, panel (c) is the 3D reconstruction for $\left|\langle\mathbf{r}|2,1,1;2^+\rangle\right|^2$.}
    \label{fig:role_of_2}
\end{figure*}

From the analytical solutions obtained for the spinor eigenstates presented in the previous section, it is straightforward to calculate the probability density distribution (for $q=1,2$), as $\rho_{nml,q^{\pm}}(\mathbf{r}) = \left|\langle \mathbf{r}|nml,q^{\pm}\rangle  \right|^2$. This function provides a clear image of the spatial localization of each spinor eigenstate within the TI-NP and, in particular, it allows us to investigate the conditions leading to the emergence of topologically trivial surface states. In Fig.~\ref{fig:role_of_n_1up}, we represent the probability density for different confined states in a TI-NP with $\alpha = \pi/6$, at the horizontal plane $z = 0.5 h$, for $q=1$, $l = m = 1$, and $1\le n \le 6$. We notice that the states with odd $n$ exhibit a finite probability density at the surface of the nanoparticle. A complementary analysis is presented in Fig.~\ref{fig:role_of_m_1up}, where the density distibution can be appreciated in 3D. Clearly, the number of maxima along the $z$-direction is determined by the corresponding quantum number $m$. In contrast, it can be shown that states with $q=2$ exhibit states at the surface for $n$ even, but with a lower probability density, as is depicted in Fig.~\ref{fig:role_of_2}.

\begin{figure*}[t!]
    \centering
    \includegraphics[scale=0.3]{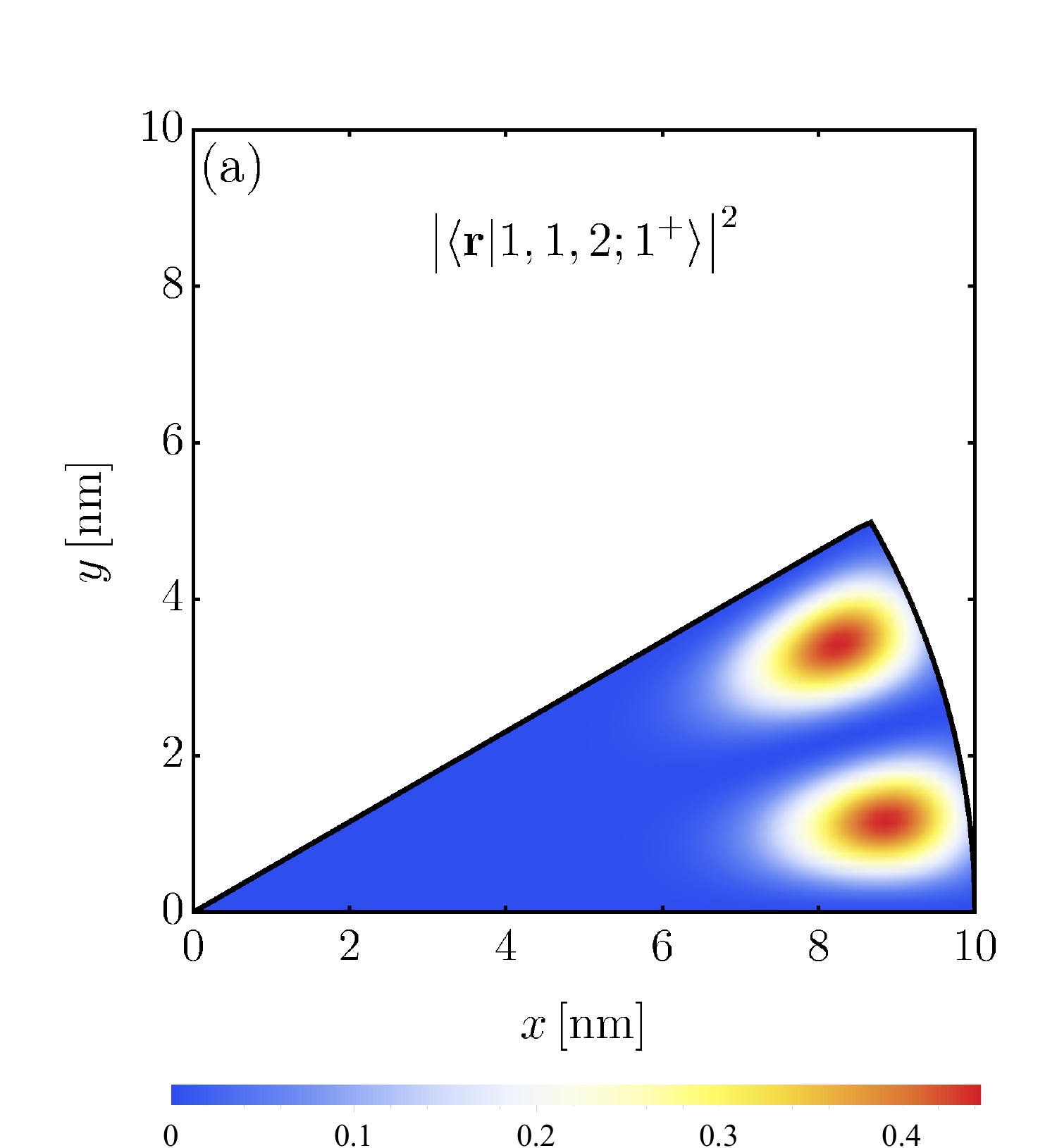}\includegraphics[scale=0.3]{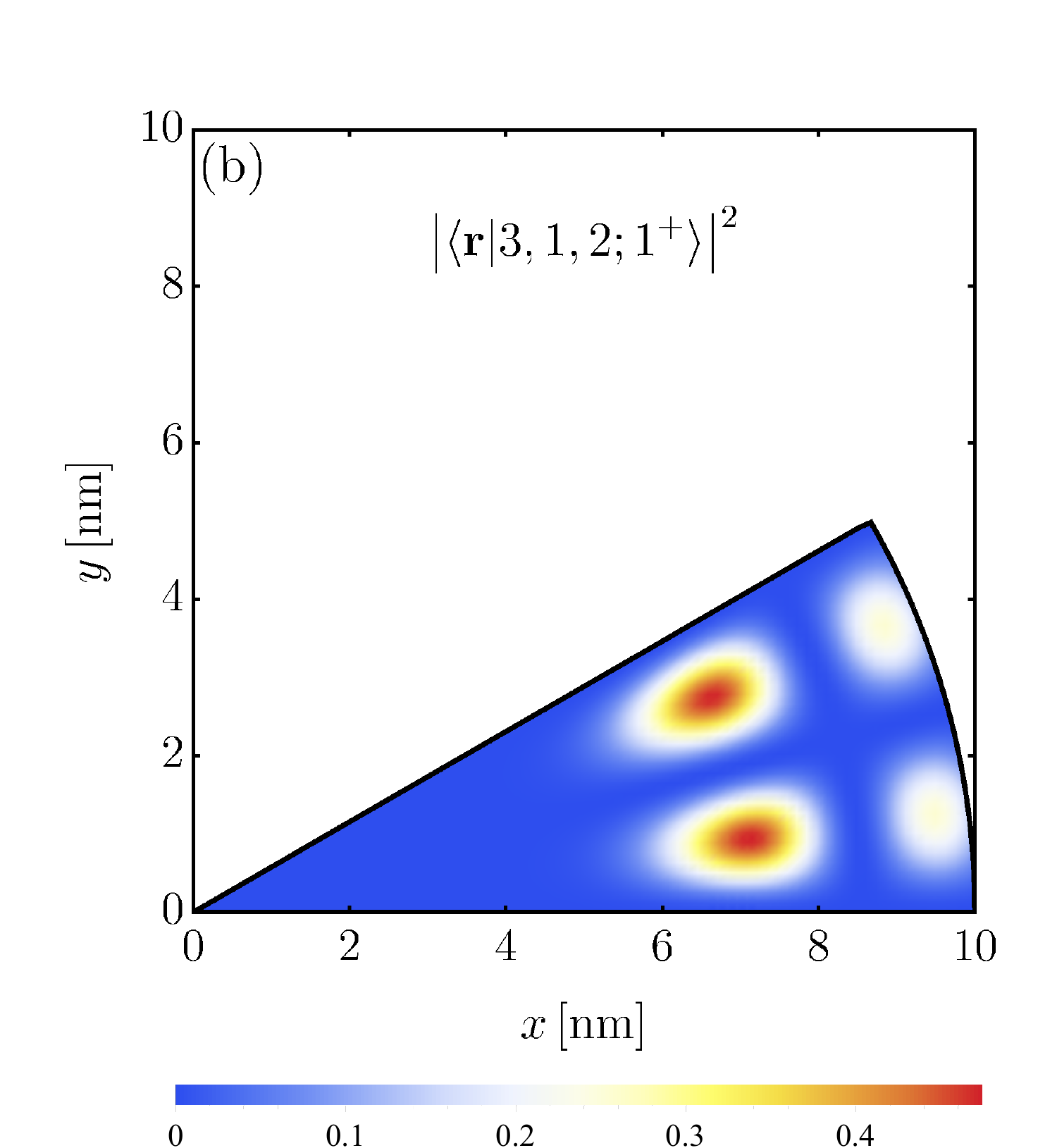}\includegraphics[scale=0.3]{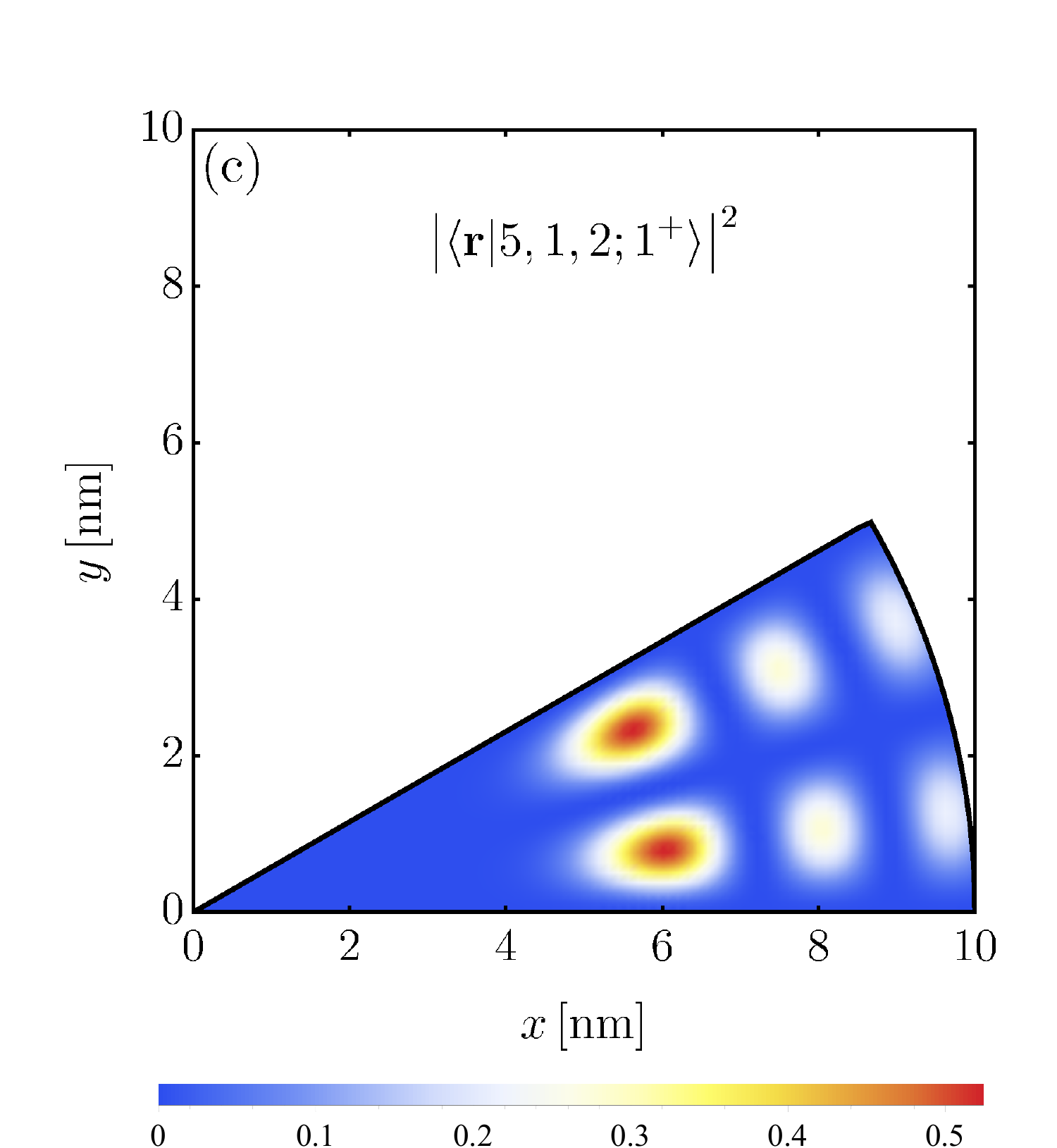}\\
    \includegraphics[scale=0.3]{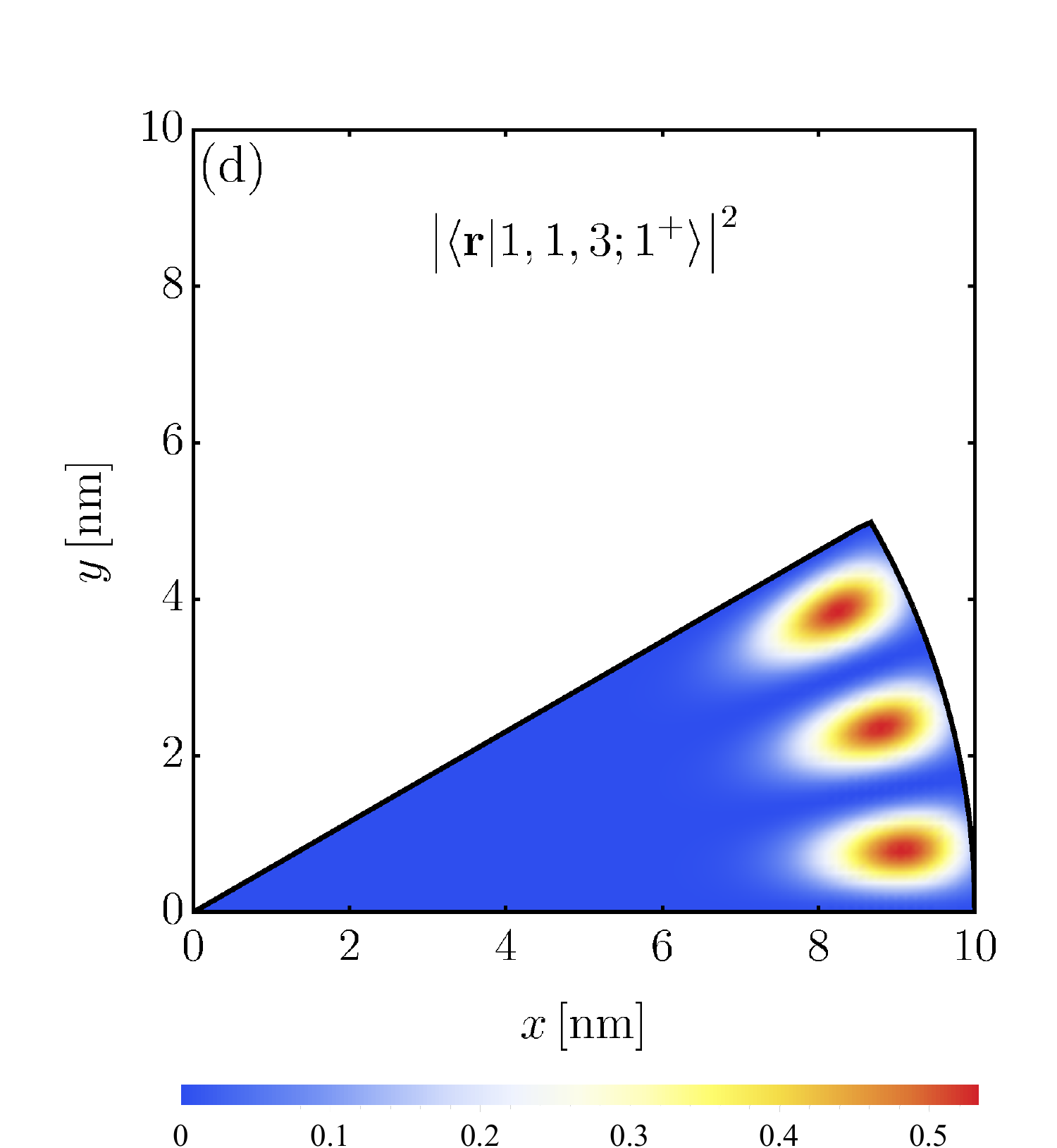}\includegraphics[scale=0.3]{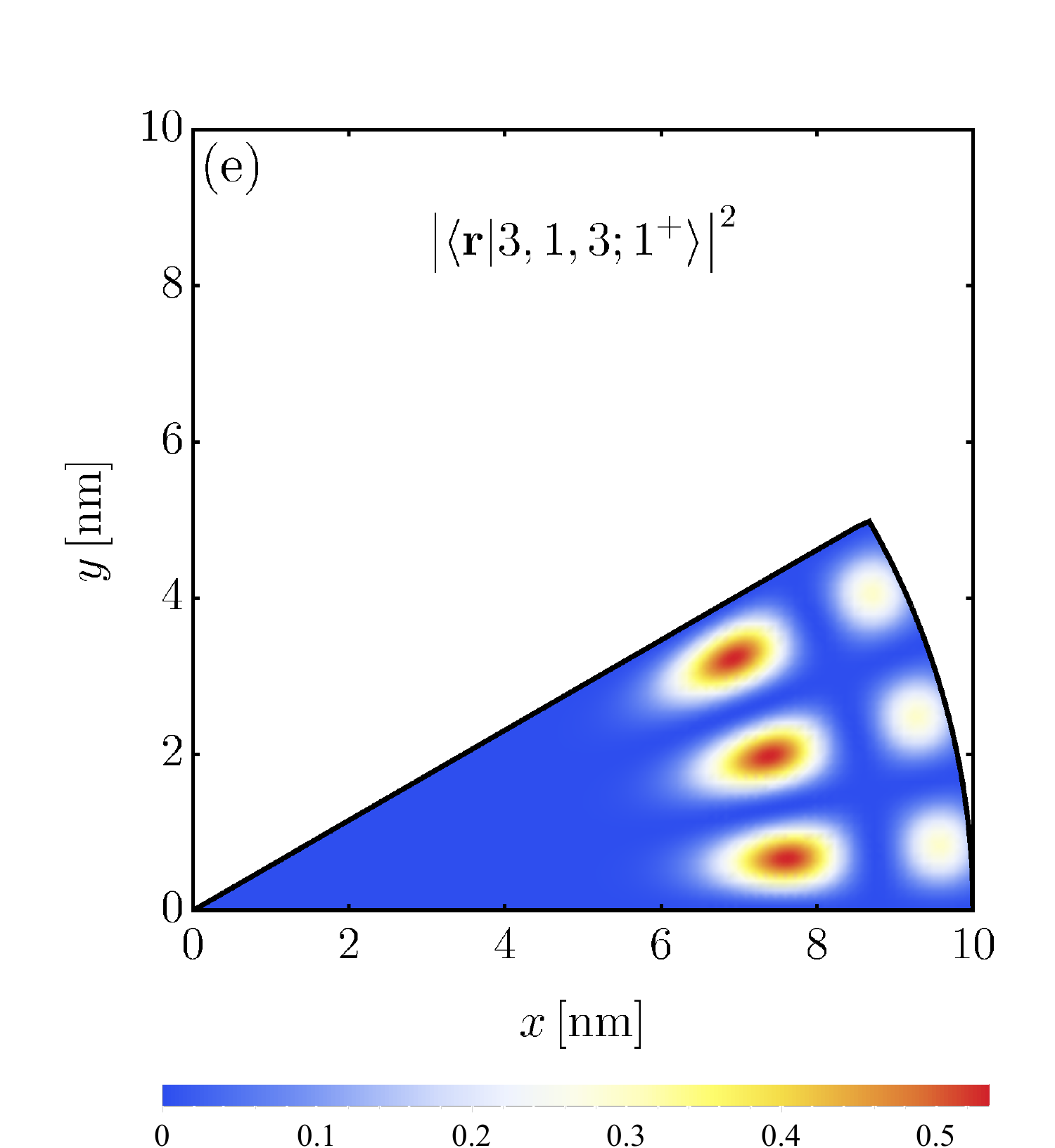}\includegraphics[scale=0.3]{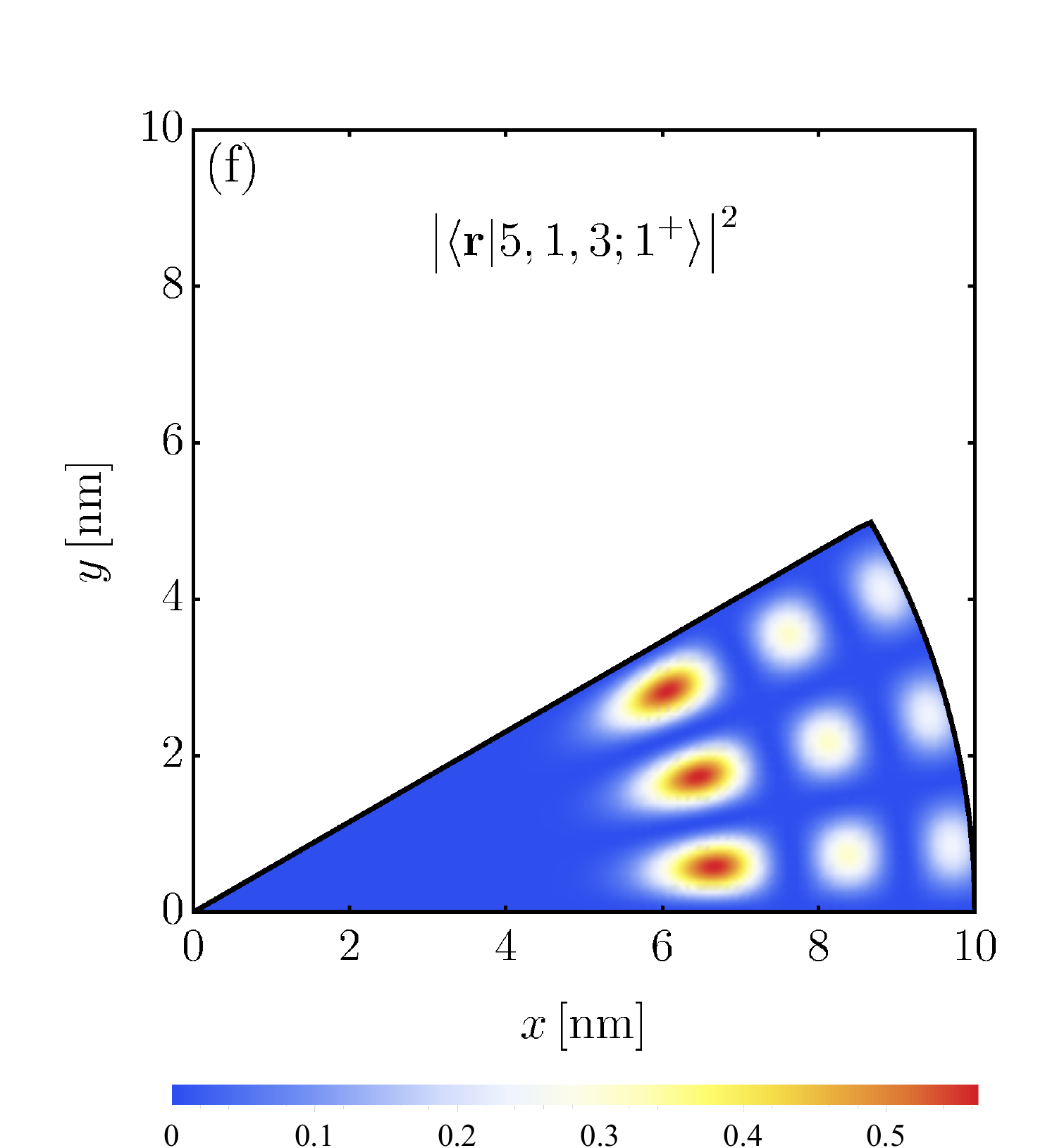}
    \caption{Probability density, represented as a 2D-color plot at the plane $z = 0.5 h$, for different confined states within a TI nanoparticle with $\alpha = \pi/6$, $m=1$, and $n = 1, 3, 5$. Subfigures (a) - (c) display $l = 2$, while Subfigures (d) - (e) correspond to $l = 3$.}
    \label{fig:role_of_l_1up}
\end{figure*}

\begin{figure*}
    \centering
    \includegraphics[scale=0.3]{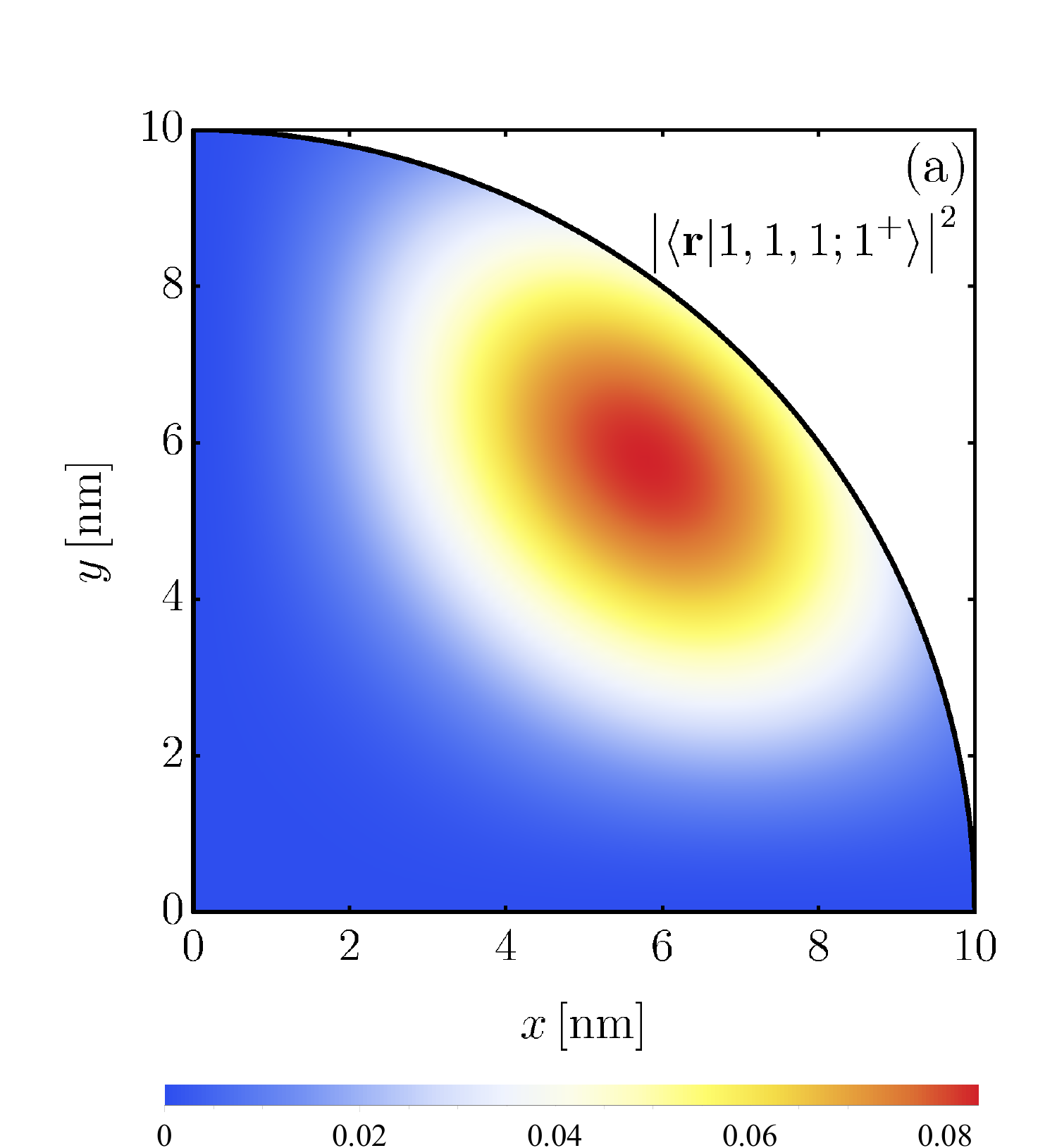}\includegraphics[scale=0.3]{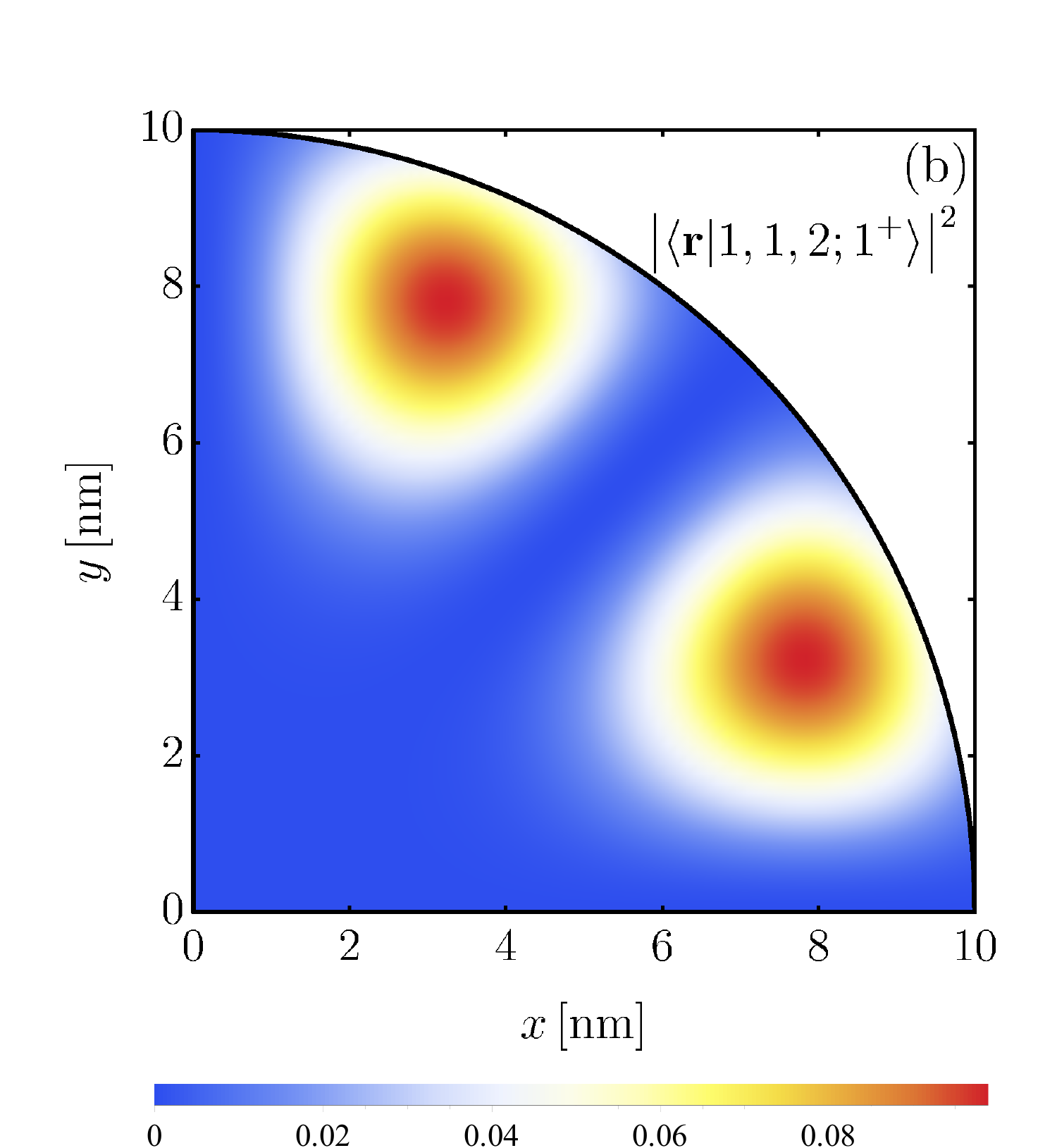}\includegraphics[scale=0.3]{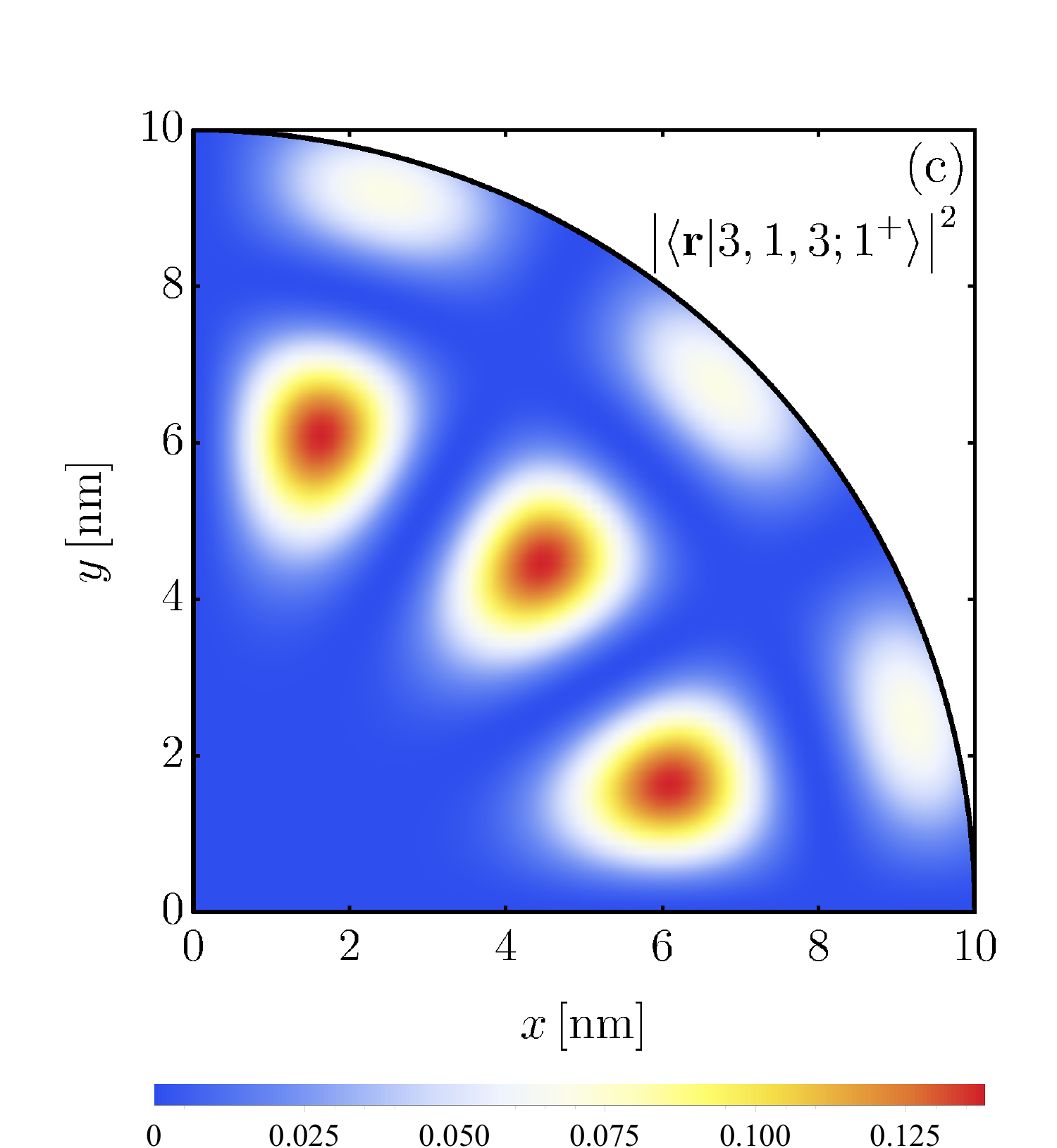}\\
    \includegraphics[scale=0.3]{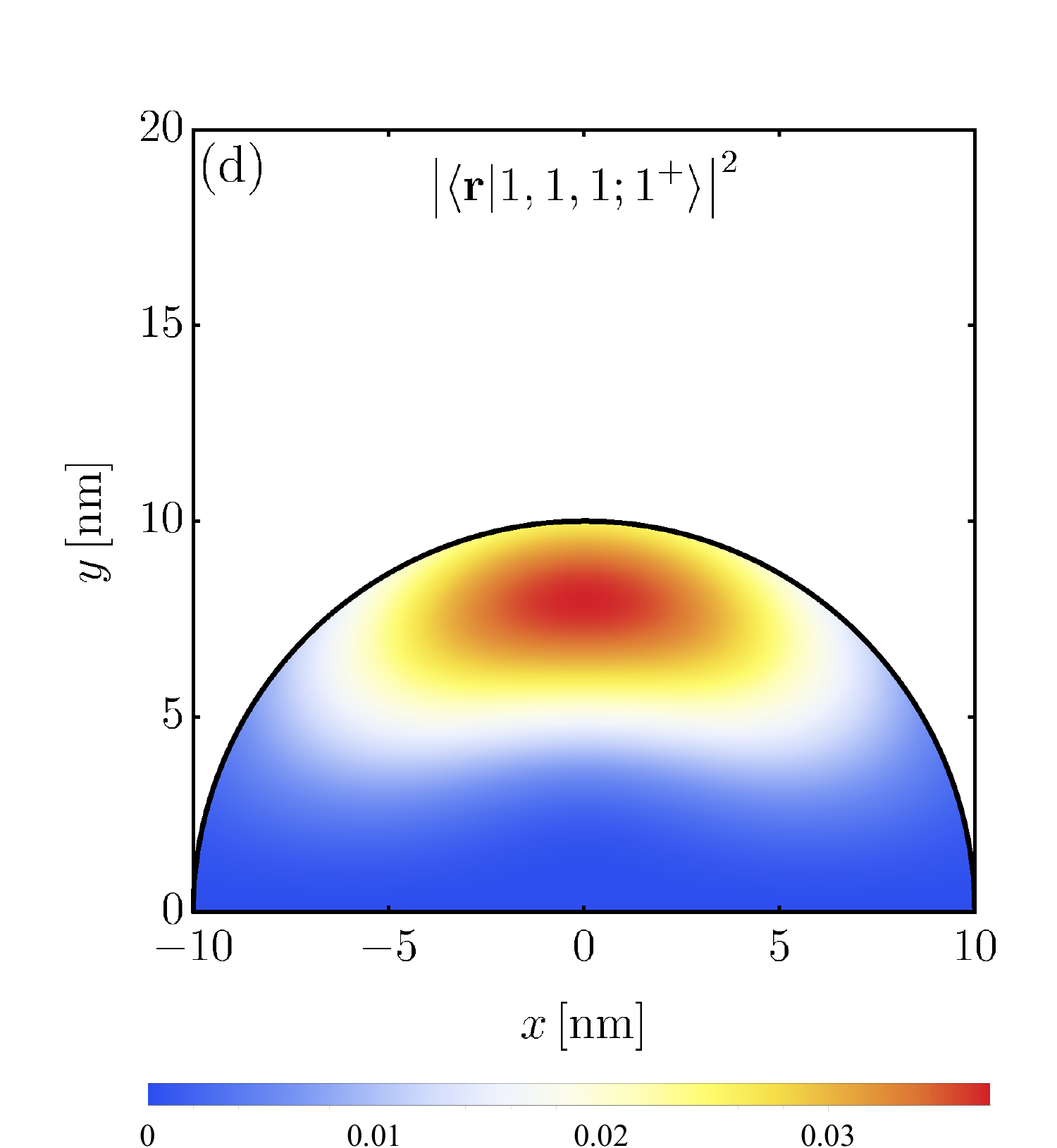}\includegraphics[scale=0.3]{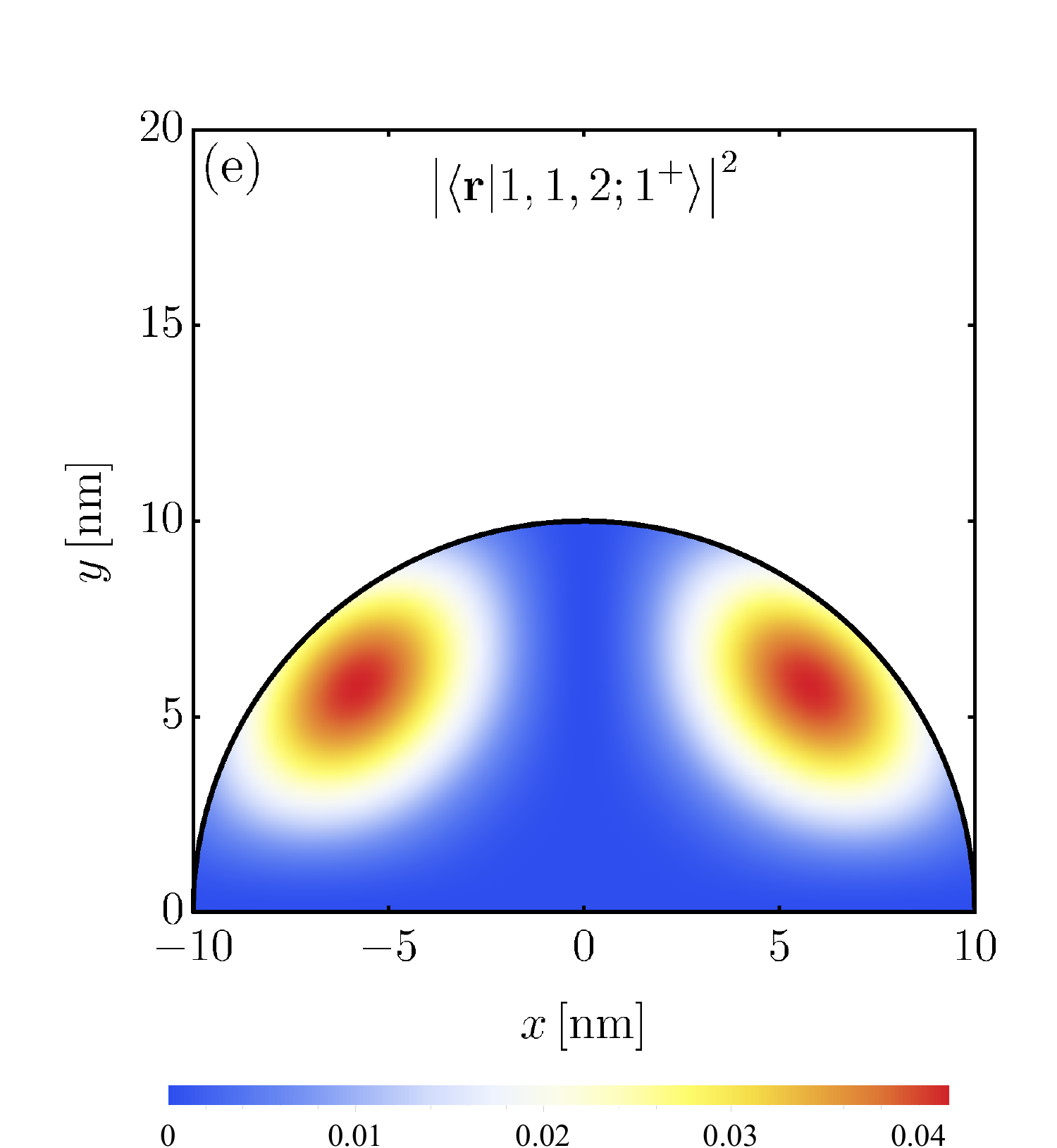}\includegraphics[scale=0.3]{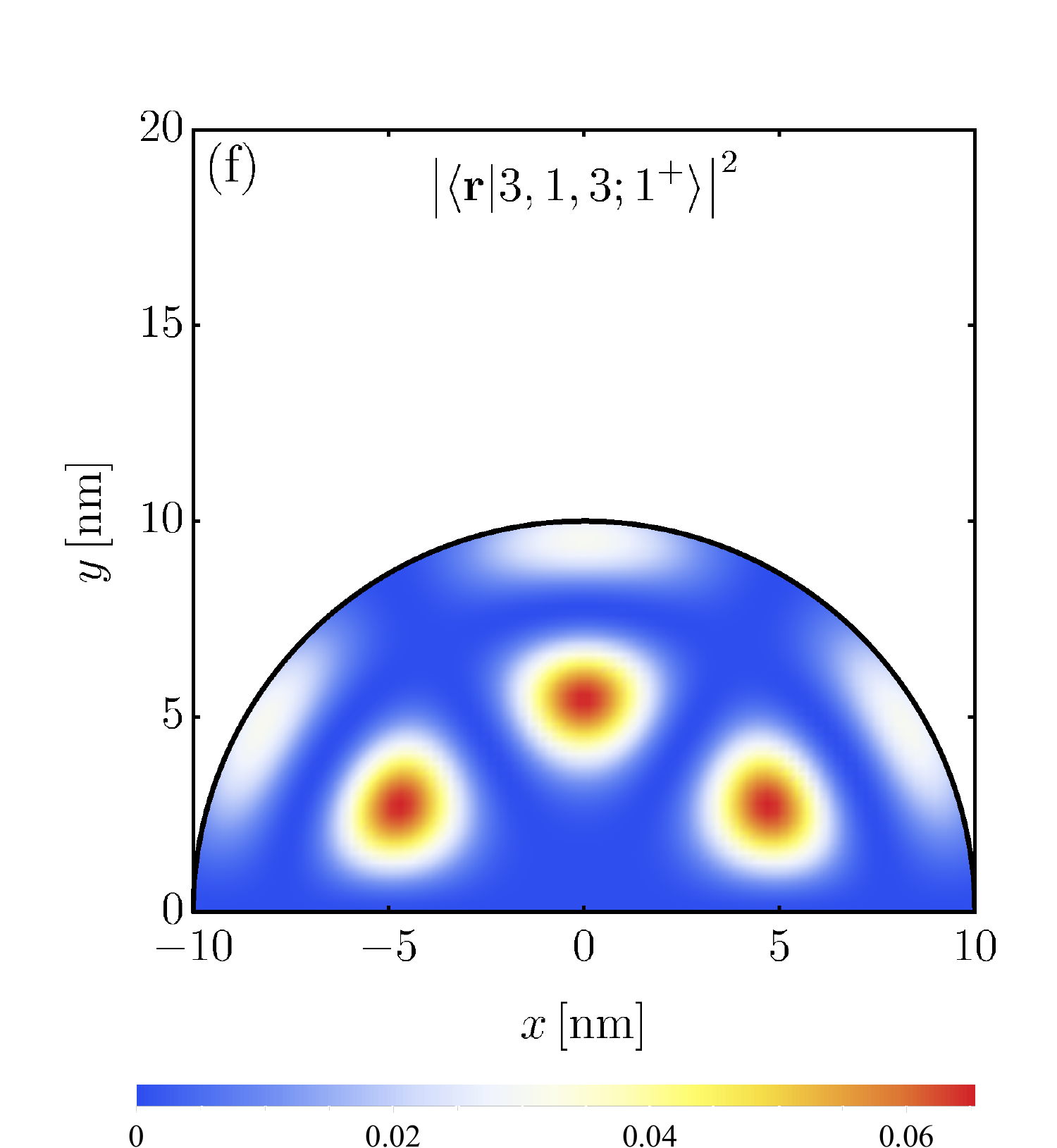}
    \caption{Probability density for $\alpha=\pi/2$ and $\alpha=\pi$ in $z=0.5h$, for different values of $n$ and $l$, and fixed $m$.}
    \label{fig:role_of_alpha_1up}
\end{figure*}

In Fig.~\ref{fig:role_of_l_1up}, we study the effect of the angular momentum quantum number $l$, characterizing the confinement effect due to the wedge angle $\alpha = \pi/6$ for this example. Its is clear from Fig.~\ref{fig:role_of_l_1up} that the number of nodes exhibited by the probability density distribution in the polar angular direction is given by the integer $l$, as seen for the examples $l=2$ in Fig.~\ref{fig:role_of_l_1up}(a-c)
and $l=3$ in Fig.~\ref{fig:role_of_l_1up}(d-f), respectively. Again we observe, in all six cases with odd $n$ ($n = 1,2,3$), the presence of a finite probability density at the surface of the TI-NP. For completeness, in order to study the impact of the geometry on the eigenstates, Fig.~\ref{fig:role_of_alpha_1up} shows the role of $\alpha$ on the spatial distribution of the probability density. We chose the angles $\alpha=\pi/2$ and $\pi$, and different values of $n$ and $l$ for $m=1$, which follows the same properties already discussed for the case $\alpha=\pi/6$.

\begin{figure*}[t!]
    \centering
    \includegraphics[scale=0.5]{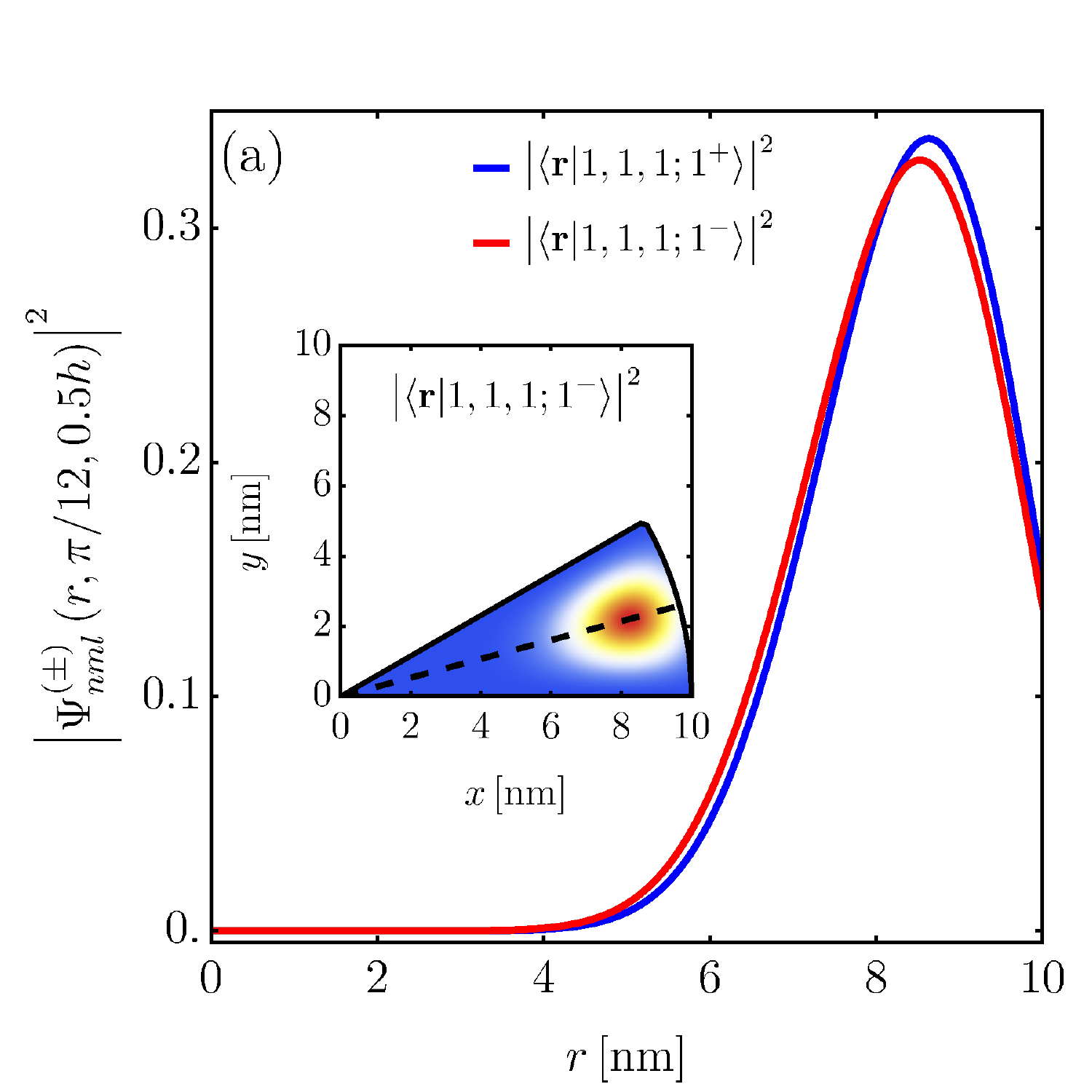}\hspace{0.5cm}\includegraphics[scale=0.5]{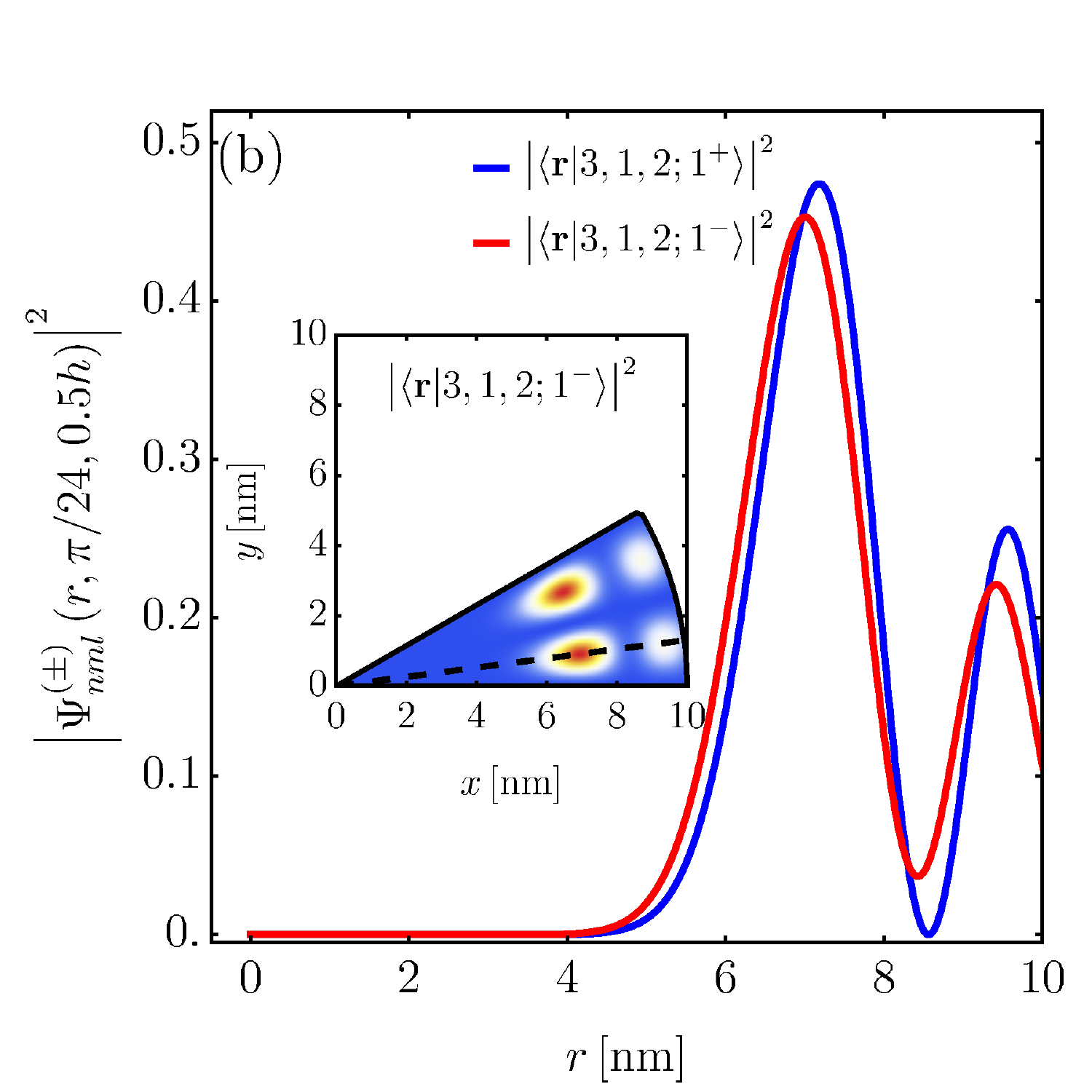}
    \caption{Comparison between the probability density distribution for the states $\ket{nml;1^+}$ and $\ket{nml;1^-}$ as a function of the radial coordinate. The insets show the 2D probability density at $z=0.5h$ and the dashed line is the direction where the main plot is constructed. }
    \label{fig:role_of_state_1}
\end{figure*}

In Fig.~\ref{fig:role_of_state_1}, we analyze the effect of the index $s=\pm$, by comparing the probability density distribution of the states $\ket{nml;1^+}$ and $\ket{nml;1^-}$, respectively, as a function of the radial coordinate at the plane $z=0.5 h$ for a TI-NP with a wedge angle $\alpha = \pi/6$. 
Fig.~\ref{fig:role_of_state_1}(a) corresponds to the state with $n=l=m=1$, along the symmetry plane $\phi=\alpha/2$ as shown in the dashed line on the inset. On the other hand, Fig.~\ref{fig:role_of_state_1}(b) corresponds to the state with $n=3$, $m=1$, and $l=2$, along the symmetry plane $\phi=\alpha/4$, as shown in the dashed line on the corresponding inset. Clearly, in both examples we observe the presence of a finite probability density at the surface of the TI-NP, but with different amplitude, given that the weight factors for each state are different.

Our analysis shows that the morphology of the TI nanoparticle as well as the values of the quantum numbers imples the existence of spatial regions with topologically trivial surface states. In general terms, depending on the parity of $n$ the quantum state $\ket{nml,q^s}$ has $m\times l$ regions in the surface $r=R$ where localized fermionic states can be found.


\section{Optical transitions and selection rules}

In this section, we analyze the optical transitions, along with the corresponding selection rules, within the standard dipolar approximation for the electric field. For this purpose, we consider the matrix elements of the total electric dipole operator~\cite{Governale_20,Gioia_19}
\bea
\mathbf{d}=e\mathbf{r}\left(\tau_0\otimes\sigma_0\right)+\frac{eR_0}{2}\left(\tau_y\otimes\boldsymbol{\sigma}\right),
\label{eq:dtot}
\eea
where the first term represents the intra-band contribution, arising from the envelope wavefunctions multiplying the same basis state in the $\mathbf{k}\cdot\mathbf{p}$ approximation for the bulk band structure~\cite{Imura_12,Governale_20,Gioia_19}. On the other hand, the second term in Eq.~\eqref{eq:dtot} accounts for transitions between different $\mathbf{k}\cdot\mathbf{p}$ basis states~\cite{Imura_12,Governale_20,Gioia_19}. 

Let us define the probability amplitude for an optical transition along the $\hat{\mathbf{e}}_{\mu}$-direction, by the matrix element
\bea
T_\mu(\ket{i}\to \ket{f})=\bra{n'm'l';f}\hat{\mathbf{e}}_{\mu}\cdot\mathbf{d}\ket{nml;i},
\label{eq:Tmu}
\eea
where $i,f$ denotes each of the four possible eigenstates labelled by the indexes $q=1,2$ and $s=\pm$, as defined in the previous section. We shall discuss our analytical results for the transition amplitudes in different orthogonal directions.

\subsection{Transition amplitudes in z-direction}
Let us first consider the transition amplitude in the z-direction, given by the analytical expression
\bea
&&T_z(\ket{1^\pm}\to \ket{1^\pm})=\frac{e\alpha}{2}A_\pm A_\pm'\delta_{ll'}\nn\\
&\times&\Bigg\{\kappa\kappa'\left(-\mathcal{M}\pm\widetilde{\mathcal{M}}\right)\left(-\mathcal{M}'\pm\widetilde{\mathcal{M}}'\right)\mathrm{I}_n^{n'}\mathrm{I}_z(\Delta m, M)\nn\\
&+&\left[kk'\left(-\mathcal{M}\pm\widetilde{\mathcal{M}}\right)\left(-\mathcal{M}'\pm\widetilde{\mathcal{M}}'\right)+1\right]\widetilde{\mathrm{I}}_{n}^{n'}\mathrm{I}_z(\Delta m, M)\nn\\
&+&\frac{\ii hR_0}{4}\delta_{mm'}\left[k'\left(-\mathcal{M}'\pm\widetilde{\mathcal{M}}'\right)-k\left(-\mathcal{M}\pm\widetilde{\mathcal{M}}\right)\right]\widetilde{\mathrm{I}}_{n}^{n'}\Bigg\},\nn\\
\eea
where the coefficients $I_n^{n'}$ and $\tilde{I}_n^{n'}$ are defined in Sec.4 of the Supplemental. In particular, for $\Delta m=m'-m$ and $M=m'+m$, the coefficient $Iz(\Delta m,M)$, defined by the analytical expression
\begin{widetext}
\bea
\mathrm{I}_z(\Delta m, M)&\equiv&\int_0^hdz~z~\sin\left(\frac{m\pi}{h}z\right)\sin\left(\frac{m'\pi}{h}z\right)\nn\\
&=&\left\{\begin{matrix}
0 & \text{ if } \Delta m\neq0 \text{ and } M\neq0\text{ are even integers.}\\
\\
4h^2/\pi^2\left(1/\Delta m^2-1/M^2\right) & \text{ If }  \Delta m \ne 0 \text{ and } M \ne 0 \text{ are odd integers} \\
\\
h^2/4 & \text{ if } \Delta m=0 \text{ and } M\neq 0 \text{ (even)} \\
\\
-h^2/4 & \text{ if }  M=0 \text{ and } \Delta m\neq 0 \text{ (even)}\\
\end{matrix}\right.,
\label{selectionrulesinM}
\eea
\end{widetext}
enforces a few basic selection rules. Moreover, we notice that in order to have a non-vanishing integral, $\Delta m$ and $M$ must have the same parity. 

As a specific numerical example, the following matrix gives the values of the first 64 elements of the transition $|T_z|$ between $\ket{111;1^+}\to\ket{nm1,1^+}$ for a TI-NP made of Bi$_2$Te$_3$ with $R=10R_0$, $h=R_0$, and $\alpha=\pi/6$:
\begin{widetext}
\bea
\begin{pmatrix}
 8 & 2.88091 & 0 & 0.23035 & 0 & 0.0634495 & 0 & 0.0261078 \\
 7.33474 & 2.64153 & 0 & 0.211215 & 0 & 0.0581791 & 0 & 0.0239392 \\
 0 & 0 & 0 & 0 & 0 & 0 & 0 & 0 \\
 1.96007 & 0.706002 & 0 & 0.0564561 & 0 & 0.015551 & 0 & 0.00639883 \\
 0 & 0 & 0 & 0 & 0 & 0 & 0 & 0 \\
 1.24078 & 0.446931 & 0 & 0.0357421 & 0 & 0.00984537 & 0 & 0.00405114 \\
 0 & 0 & 0 & 0 & 0 & 0 & 0 & 0 \\
 0.936291 & 0.337241 & 0 & 0.026972 & 0 & 0.00742969 & 0 & 0.00305715 \\
\end{pmatrix}
\eea
\end{widetext}

\begin{widetext}
Analogous analytical expressions are obtained for the following transition amplitudes:
\bea
T_z(\ket{2^\pm}\to \ket{2^\pm})
&=&\frac{e\alpha}{2}B_\pm B_\pm'\delta_{ll'}\Bigg\{\kappa\kappa'\left(\mathcal{M}\pm\widetilde{\mathcal{M}}\right)\left(\mathcal{M}'\pm\widetilde{\mathcal{M}}'\right)\widetilde{\mathrm{I}}_n^{n'}\mathrm{I}_z(\Delta m, M)\nn\\
&+&\left[kk'\left(\mathcal{M}\pm\widetilde{\mathcal{M}}\right)\left(\mathcal{M}'\pm\widetilde{\mathcal{M}}'\right)+1\right]\mathrm{I}_{n}^{n'}\mathrm{I}_z(\Delta m, M)\nn\\
&+&\frac{\ii hR_0}{4}\delta_{mm'}\left[k\left(\mathcal{M}\pm\widetilde{\mathcal{M}}\right)-k'\left(\mathcal{M}'\pm\widetilde{\mathcal{M}}'\right)\right]\mathrm{I}_{n}^{n'}\Bigg\},
\eea

\bea
T_z(\ket{1^\pm}\to \ket{1^\mp})
&=&\frac{e\alpha}{2}A_\pm A_\mp'\delta_{ll'}\Bigg\{\kappa\kappa'\left(-\mathcal{M}\pm\widetilde{\mathcal{M}}\right)\left(-\mathcal{M}'\mp\widetilde{\mathcal{M}}'\right)\mathrm{I}_n^{n'}\mathrm{I}_z(\Delta m, M)\nn\\
&+&\left[kk'\left(-\mathcal{M}\pm\widetilde{\mathcal{M}}\right)\left(-\mathcal{M}'\mp\widetilde{\mathcal{M}}'\right)+1\right]\widetilde{\mathrm{I}}_{n}^{n'}\mathrm{I}_z(\Delta m, M)\nn\\
&+&\frac{\ii hR_0}{4}\delta_{mm'}\left[k'\left(-\mathcal{M}'\mp\widetilde{\mathcal{M}}'\right)-k\left(-\mathcal{M}\pm\widetilde{\mathcal{M}}\right)\right]\widetilde{\mathrm{I}}_{n}^{n'}\Bigg\},
\eea

\bea
T_z(\ket{2^\pm}\to \ket{2^\mp})
&=&\frac{e\alpha}{2}B_\pm B_\mp'\delta_{ll'}\Bigg\{\kappa\kappa'\left(\mathcal{M}\pm\widetilde{\mathcal{M}}\right)\left(\mathcal{M}'\mp\widetilde{\mathcal{M}}'\right)\widetilde{\mathrm{I}}_n^{n'}\mathrm{I}_z(\Delta m, M)\nn\\
&+&\left[kk'\left(\mathcal{M}\pm\widetilde{\mathcal{M}}\right)\left(\mathcal{M}'\mp\widetilde{\mathcal{M}}'\right)+1\right]\mathrm{I}_{n}^{n'}\mathrm{I}_z(\Delta m, M)\nn\\
&+&\frac{\ii hR_0}{4}\delta_{mm'}\left[k\left(\mathcal{M}\pm\widetilde{\mathcal{M}}\right)-k'\left(\mathcal{M}'\mp\widetilde{\mathcal{M}}'\right)\right]\mathrm{I}_{n}^{n'}\Bigg\},
\eea

\bea
T_z(\ket{1^\pm}\to \ket{2^\pm})
&=&\frac{e\alpha}{2}A_\pm B_\pm'\delta_{ll'}\Bigg\{\ii \left(\mathcal{M}'\pm\widetilde{\mathcal{M}}'\right)\left(-\mathcal{M}\pm\widetilde{\mathcal{M}}\right)\left[k'\kappa~\mathrm{I}_n^{n'}-k\kappa'~\widetilde{\mathrm{I}}_n^{n'}\right]\mathrm{I}_z(\Delta m, M)\nn\\
&+&\frac{hR_0}{4}\delta_{mm'}\left[\kappa'\left(\mathcal{M}'\pm\widetilde{\mathcal{M}}'\right)\widetilde{\mathrm{I}}_n^{n'}-\kappa\left(-\mathcal{M}\pm\widetilde{\mathcal{M}}\right)\mathrm{I}_n^{n'}\right]\Bigg\},
\eea
and
\bea
T_z(\ket{1^\pm}\to \ket{2^\mp})
&=&\frac{e\alpha}{2}A_\pm B_\pm'\delta_{ll'}\Bigg\{\ii \left(\mathcal{M}'\mp\widetilde{\mathcal{M}}'\right)\left(-\mathcal{M}\pm\widetilde{\mathcal{M}}\right)\left[k'\kappa~\mathrm{I}_n^{n'}-k\kappa'~\widetilde{\mathrm{I}}_n^{n'}\right]\mathrm{I}_z(\Delta m, M)\nn\\
&+&\frac{hR_0}{4}\delta_{mm'}\left[\kappa'\left(\mathcal{M}'\mp\widetilde{\mathcal{M}}'\right)\widetilde{\mathrm{I}}_n^{n'}-\kappa\left(-\mathcal{M}\pm\widetilde{\mathcal{M}}\right)\mathrm{I}_n^{n'}\right]\Bigg\}.
\eea
\end{widetext}

\begin{figure}[h!]
    \centering
    \includegraphics[scale=0.5]{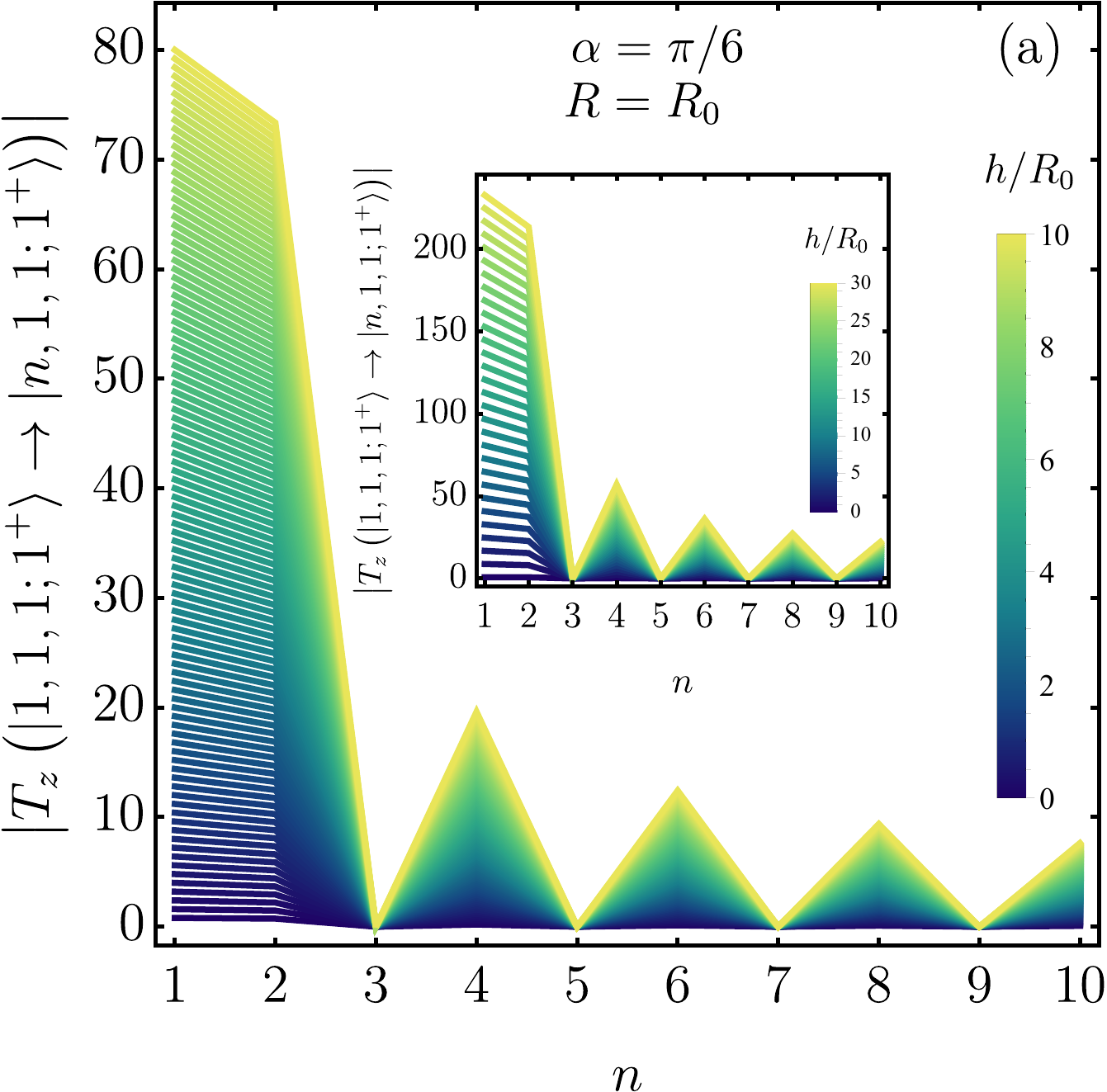}\\\vspace{0.5cm}\includegraphics[scale=0.5]{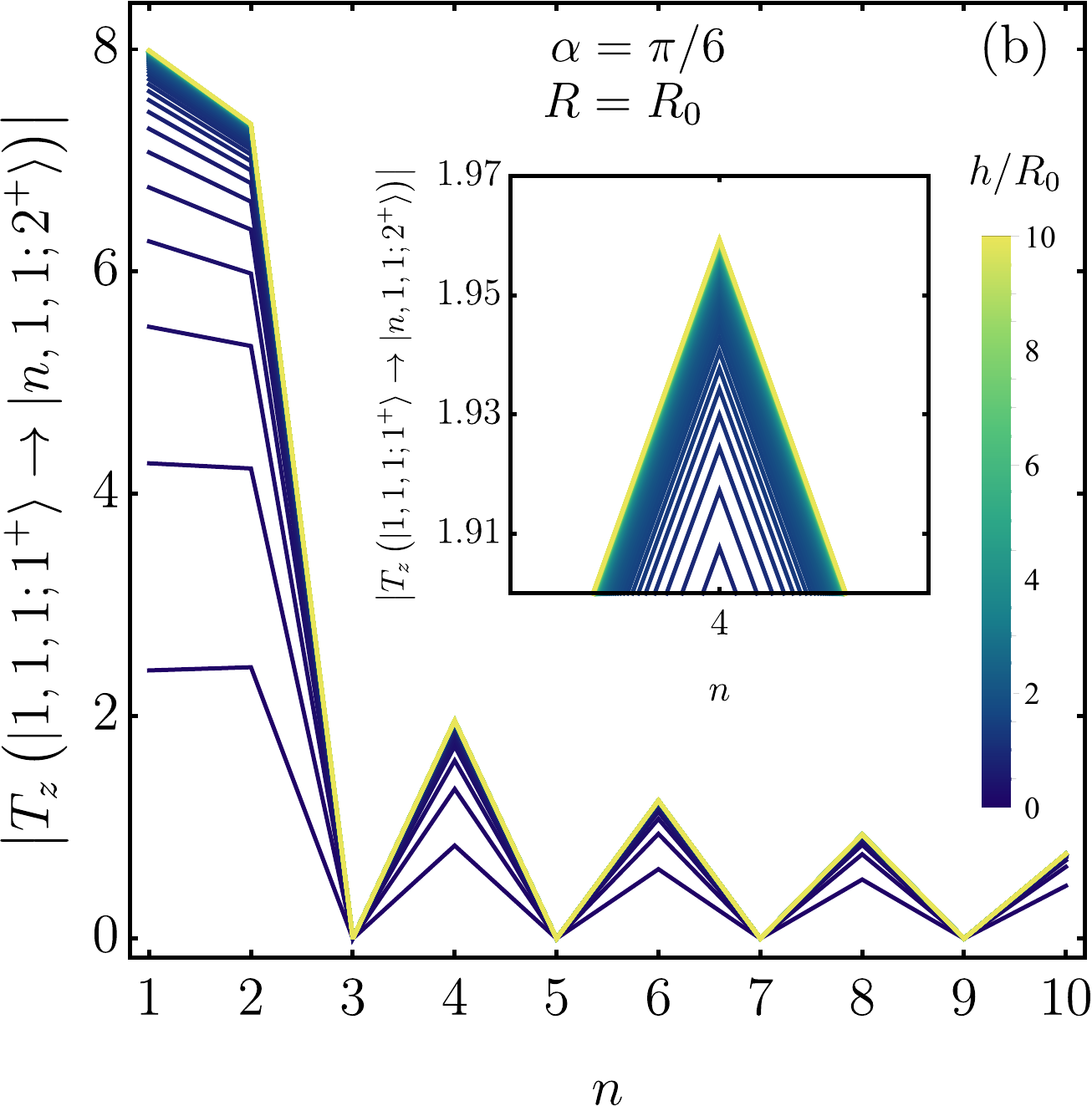}
    \caption{Selection rule in $z$-direction for a TI nanoparticle made of Bi$_2$Se$_3$ with a radius $R=R_0$ and $\alpha=\pi/6$. The figure shows the allowed and forbidden transitions from the ground state as a function of the height $h$.}
    \label{fig:Tz}
\end{figure}

Figure~\ref{fig:Tz} shows the selection rules in Bi$_2$Se$_3$ for the transition $\ket{1,1,1;1^+}\to\ket{n,1,1;1^+}$ (Fig.~\ref{fig:Tz}-(a)), and $\ket{1,1,1;1^+}\to\ket{n,1,1;2^+}$ (Fig.~\ref{fig:Tz}-(b)) and  as a function of $n$ and the ratio $h/R_0$ for $\alpha=\pi/6$ and $R=R_0$, respectively. In addition to the conditions stated in Eq.~(\ref{selectionrulesinM}) and $l=l'$, it is clear that transitions with $\Delta n$ even are forbidden. Moreover, we find that the amplitude of the matrix elements for $T_z$ increases proportionally to $h/R_0$ for $\ket{1,1,1;1^+}\to\ket{n,1,1;1^+}$, as can be noticed in the inset of Fig.~\ref{fig:Tz}-(a). This finite-size effect is related to the delocalization of the particle in the $z$-direction. In fact, the element $\ket{1,1,1;1^+}\to\ket{1,1,1;1^+}$ can be interpreted as a polarization of the electronic wave function, which spreads over all the region when $h\to\infty$ (it turns to be a plane wave $e^{\pm\ii k z}$ where the dipole approximation is not longer valid). On the other hand, the case $\ket{1,1,1;1^+}\to\ket{n,1,1;2^+}$ presents a saturation value displayed by the yellow line in Fig.~\ref{fig:Tz}-(b). Finally, our numerical calculations demonstrate that for these transition amplitudes the values of $R$ and $\alpha$ do not provide an appreciable change in the profiles of Fig.~\ref{fig:Tz}. 

\subsection{Transition amplitudes in $x\pm\ii y$-direction}
Let us now consider the transition amplitudes in the $x\pm\ii y$-direction, given by the analytical expression
\bea
&&T_{x\pm\ii y}(\ket{1^\pm}\to \ket{1^\pm})
=\frac{eh}{2}A_\pm A_\pm'\delta_{mm'}\nn\\
&\times&\Bigg\{\kappa\kappa'\left(-\mathcal{M}\pm\widetilde{\mathcal{M}}\right)\left(-\mathcal{M}'\pm\widetilde{\mathcal{M}}'\right)\mathrm{I}_\phi^{\pm}(\Delta l,L)\mathrm{J}_{nl}^{n'l'}\nn\\
&+&\left[kk'\left(-\mathcal{M}\pm\widetilde{\mathcal{M}}\right)\left(-\mathcal{M}'\pm\widetilde{\mathcal{M}}'\right)+1\right]\mathrm{I}_\phi^{\pm}(\Delta l,L)\widetilde{\mathrm{J}}_{nl}^{n'l'}\nn\\
&-&\ii R_0\kappa' \left(-\mathcal{M}'\pm\widetilde{\mathcal{M}}'\right)\mathrm{I}_\phi^{+}(\Delta l,L)\widetilde{\mathrm{K}}_{nl}^{n'l'}\Bigg\},
\label{Txy1to1}
\eea
where the coefficients $J_{nl}^{n'l'}$, $\tilde{J}_{nl}^{n'l'}$ and $\tilde{K}_{nl}^{n'l'}$ are defined in Sec.~3 of the Supplemental. On the other hand, for $\Delta l=l'-l$ and $L=l+l'$, the coefficients $\mathrm{I}_\phi^{\pm}(\Delta l,L)$ are defined by the expression
\bea
&&\mathrm{I}_\phi^{\pm}(\Delta l,L)\equiv\int_0^\alpha d\phi~e^{\pm \ii\phi}\sin\left(\frac{l\pi}{\alpha}\phi\right)\sin\left(\frac{l'\pi}{\alpha}\phi\right)\nn\\
&=&\frac{\alpha}{2}\frac{e^{\pm\ii\alpha}\left[\pi\Delta l\sin(\pi\Delta l)\pm\ii\alpha\cos(\pi\Delta l)\right]\mp\ii\alpha}{\left(\pi^2\Delta l^2-\alpha^2\right)}\nn\\
&-&\frac{\alpha}{2}\frac{e^{\pm\ii\alpha}\left[\pi L\sin(\pi L)\pm\ii\alpha\cos(\pi L)\right]\mp\ii\alpha}{\left(\pi^2L^2-\alpha^2\right)}.
\label{eq:Iphi}
\eea
\begin{figure}[h!]
    \centering
    \includegraphics[scale=0.5]{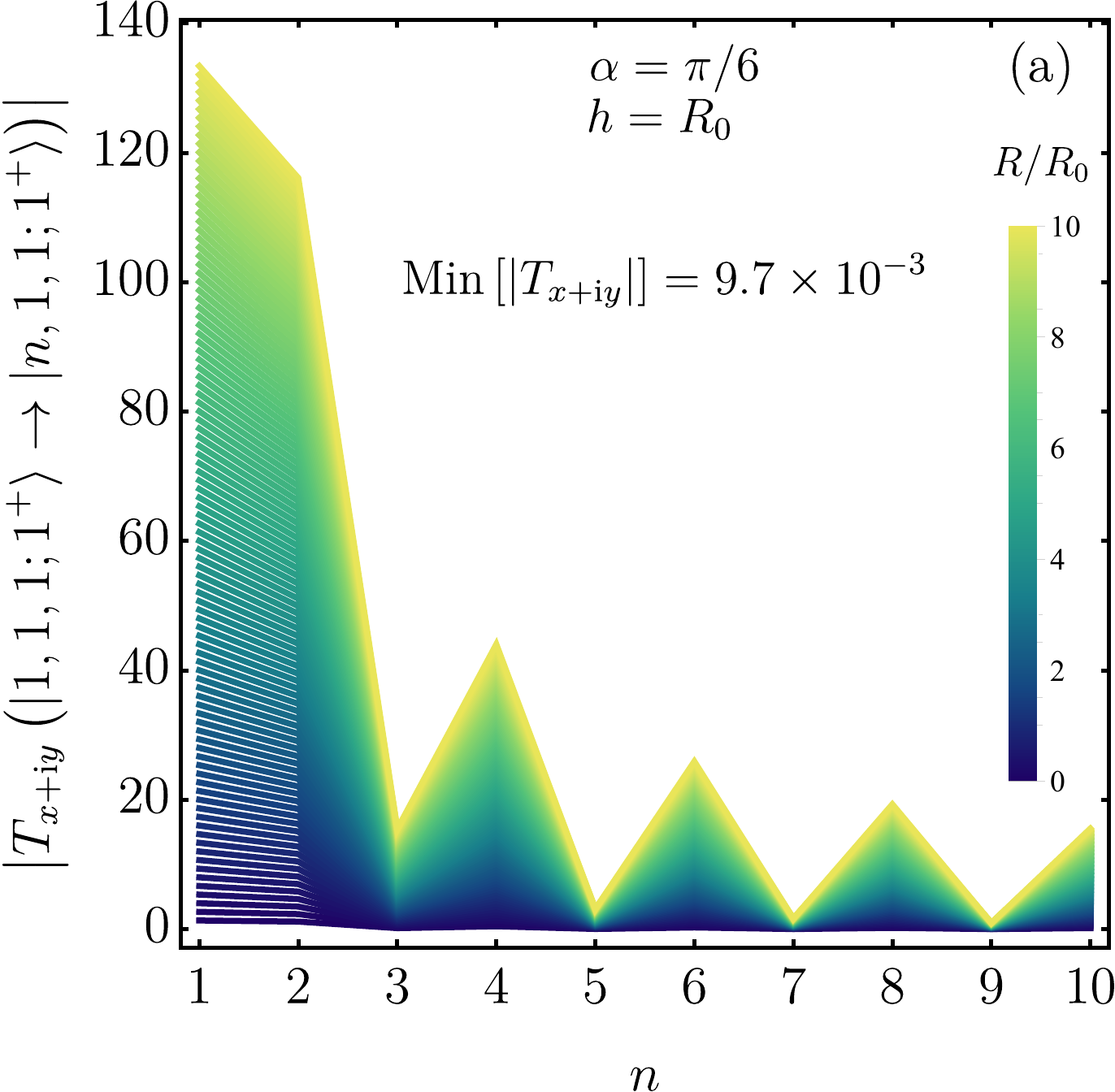}\\\vspace{0.5cm}\includegraphics[scale=0.5]{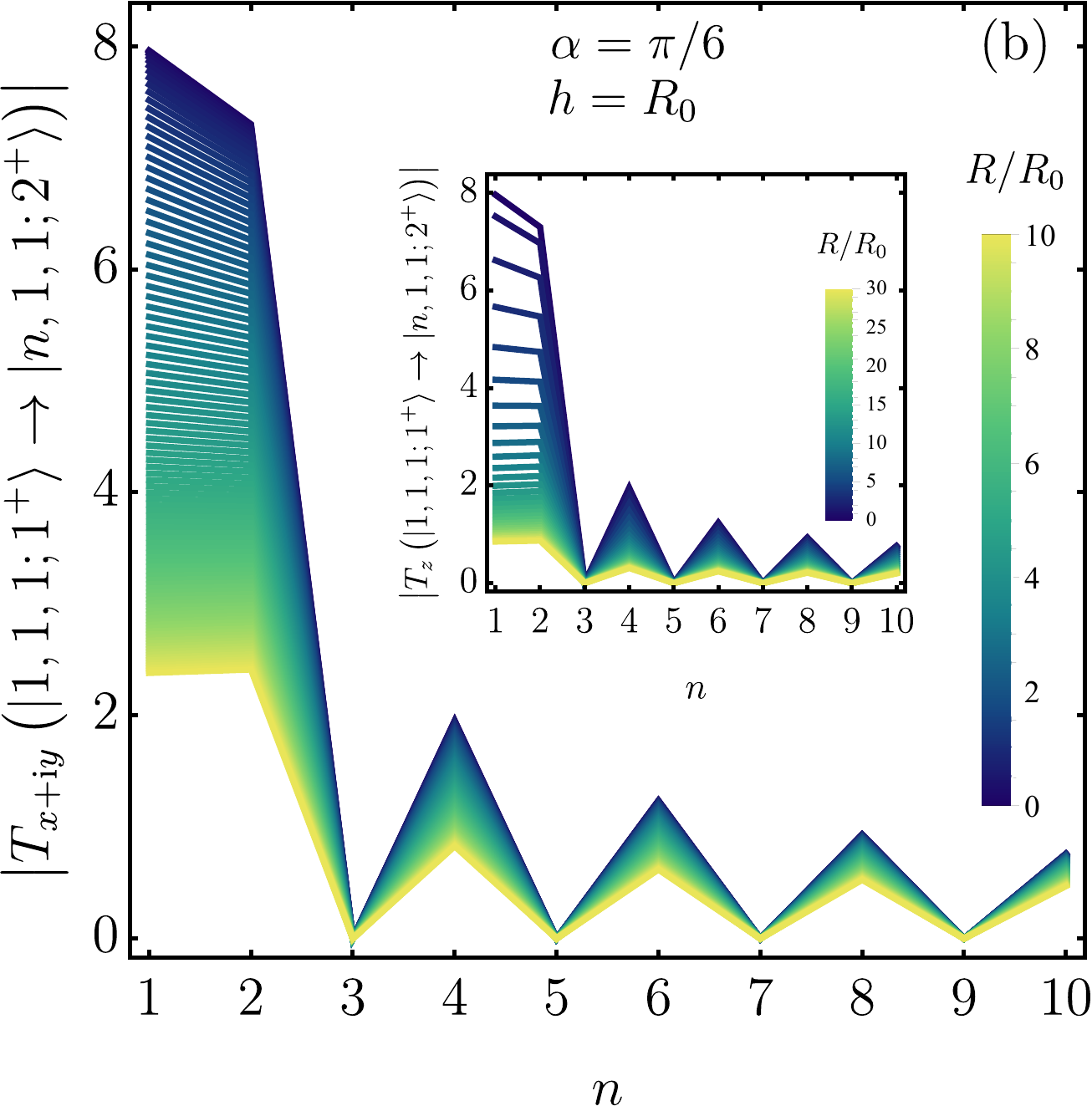}
    \caption{Selection rule in $x+\ii y$-direction for a TI nanoparticle made of Bi$_2$Se$_3$ with a height $h=R_0$ and $\alpha=\pi/6$. The figure shows the allowed and forbidden transitions from the ground state as a function of the radius $R$.}
    \label{fig:Txy}
\end{figure}

\begin{figure*}[t!]
    \centering
    \includegraphics[scale=0.3]{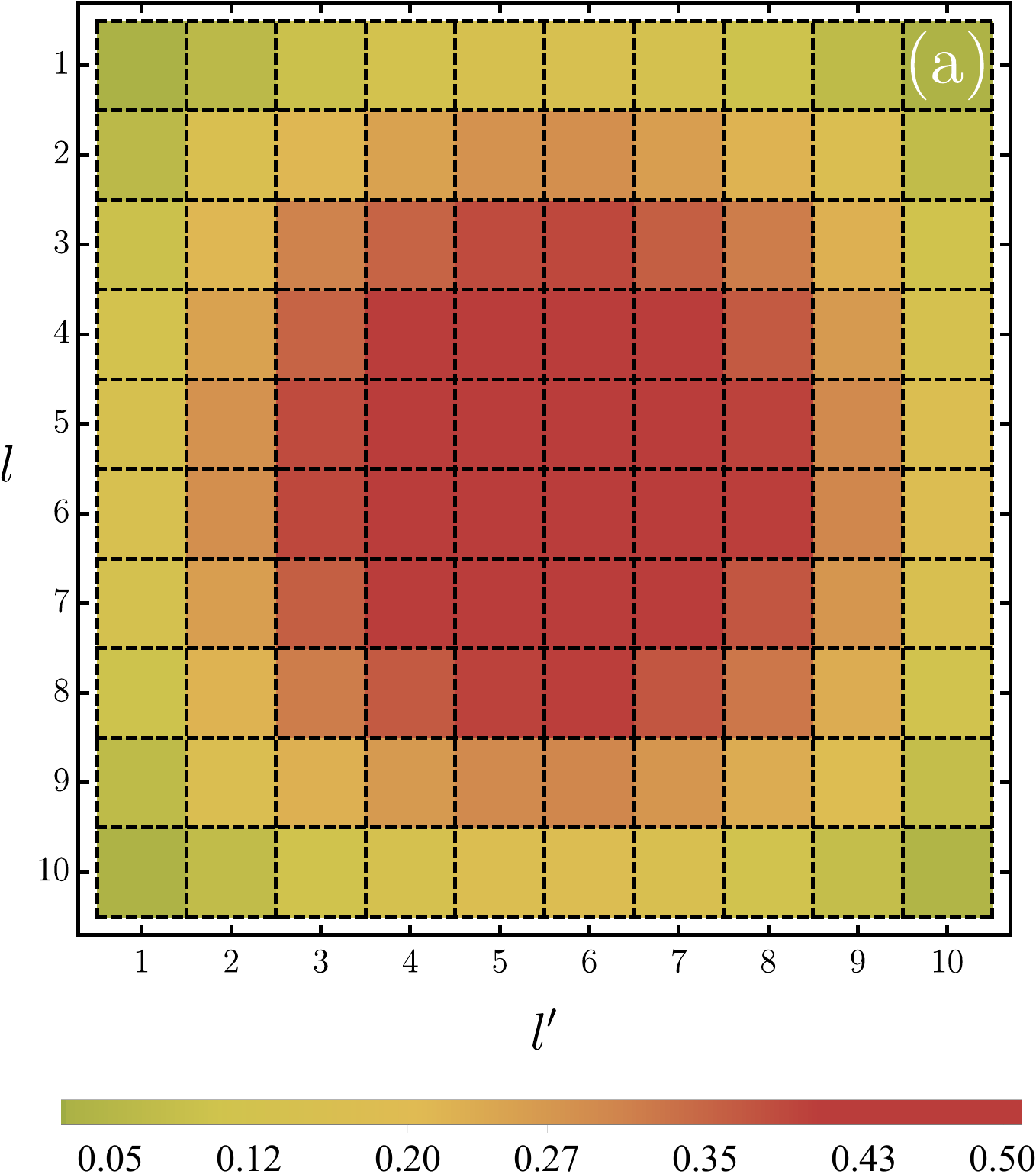}\hspace{0.2cm}\includegraphics[scale=0.3]{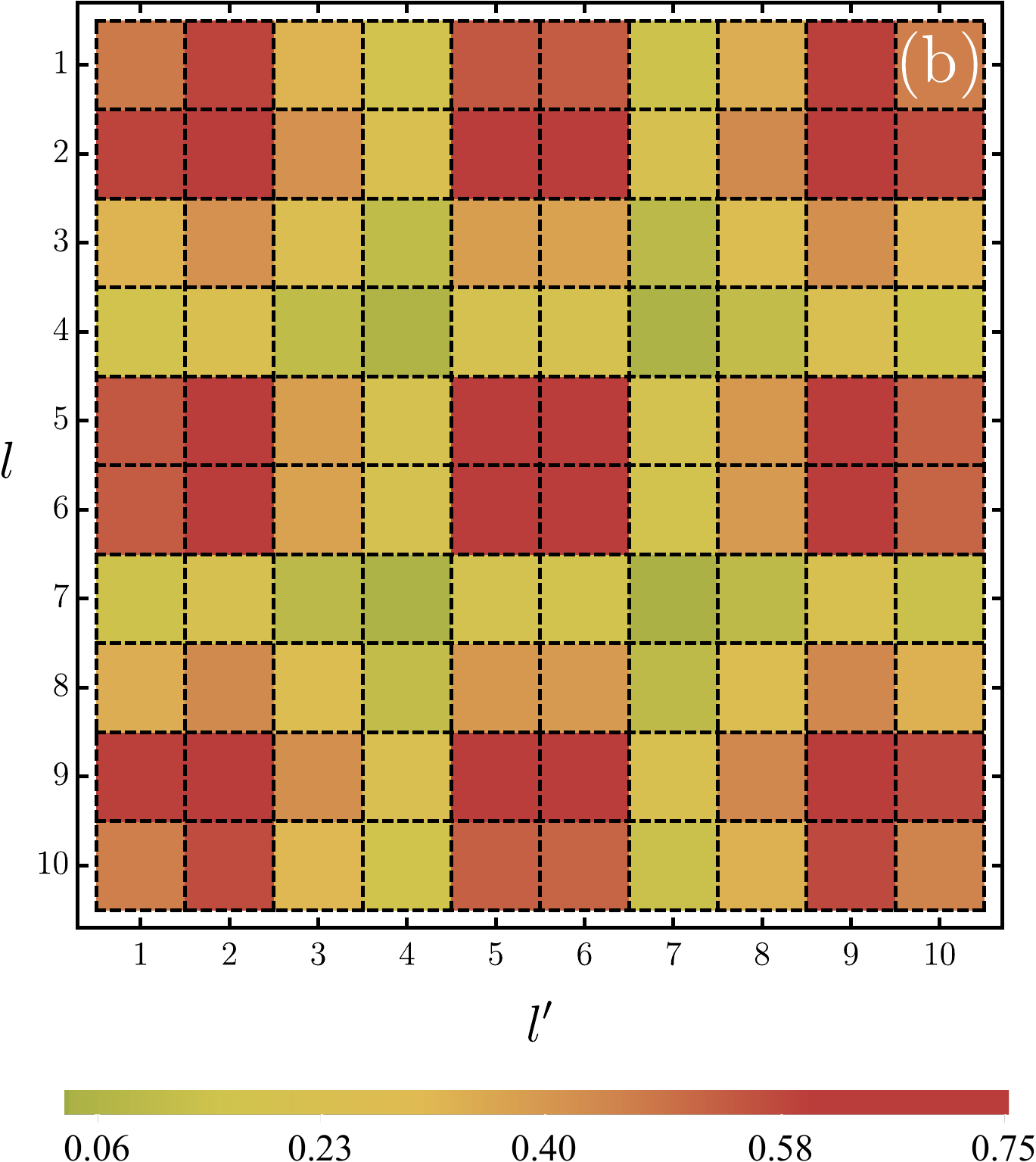}\hspace{0.2cm}\includegraphics[scale=0.3]{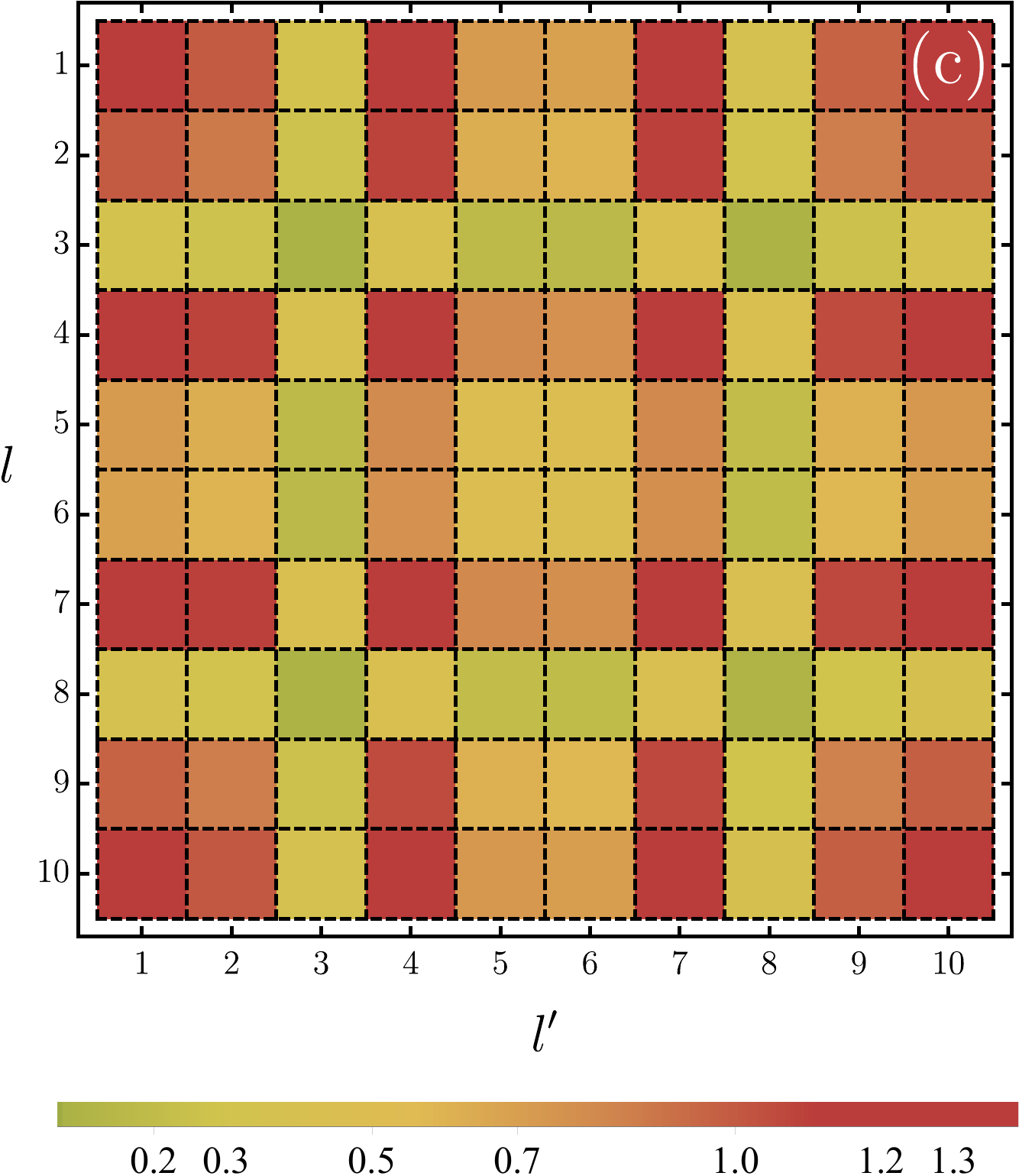}\hspace{0.2cm}\includegraphics[scale=0.3]{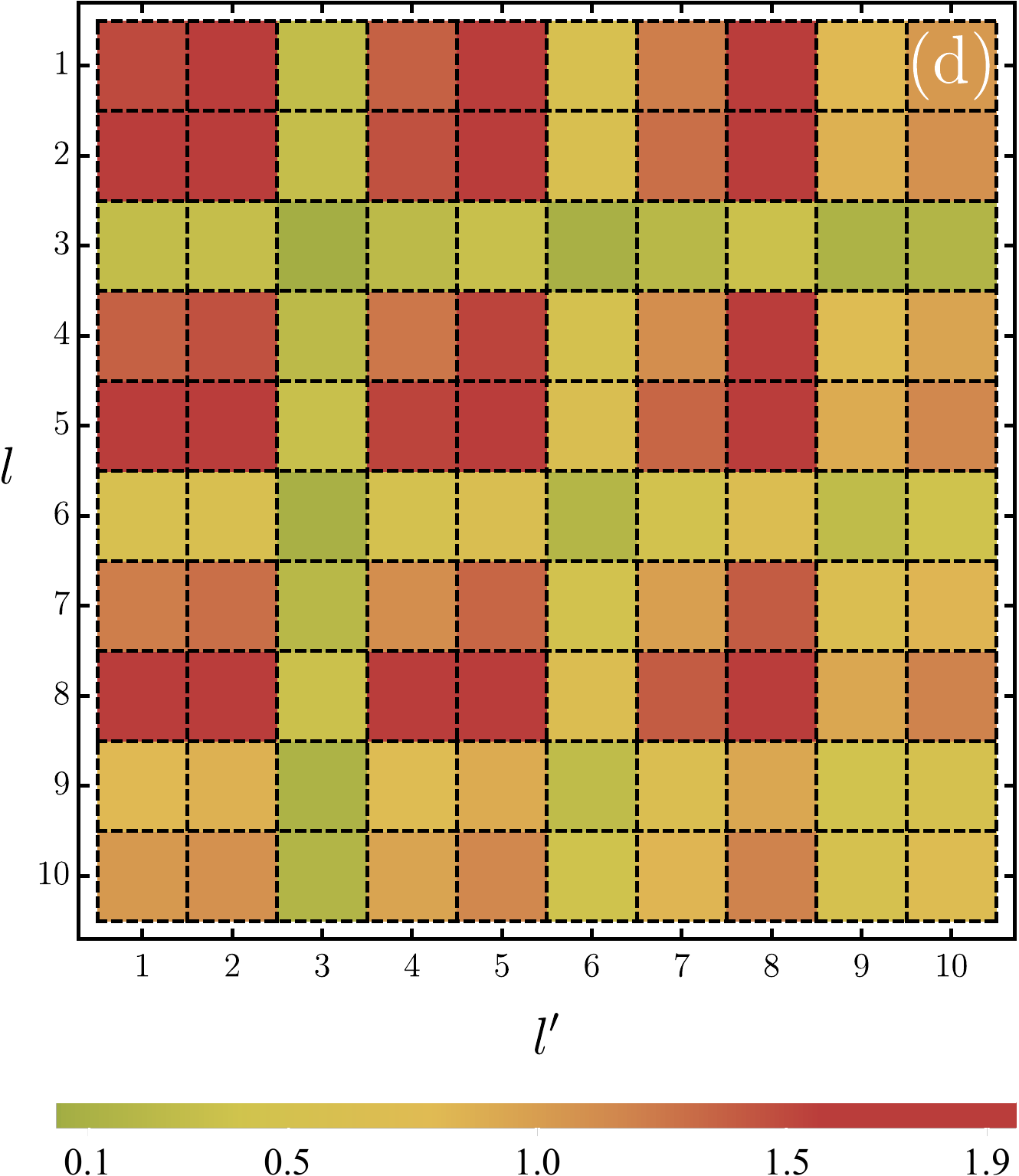}
    \caption{Tomography of the integral $\left|\mathrm{I}_\phi^+\right|$ defined in Eq.~\eqref{eq:Iphi} as a function of the angular quantum number for the initial ($l$) and final ($l'$) states. Panel (a) is the case for $\alpha=\pi/6$, panel (b) for $\alpha=\pi/4$, (c) for $\alpha=\pi/2$, and (d) for $\alpha=\pi$.}
    \label{fig:Iphi}
\end{figure*}
 The selection rules in the $x+\ii y$-direction are understood from the information presented in Figs.~\ref{fig:Txy} and~\ref{fig:Iphi}. In this case, the relevant confinement parameter is $R$, so that variations on $h$ and $\alpha$ produce only tiny effects. Also, the transition $\ket{1,1,1;1^+}\to\ket{n,1,1;1^+}$ is enhanced as $R$ grows, while the transition $\ket{1,1,1;1^+}\to\ket{n,1,1;2^+}$ is reduced. Moreover, in Fig.~\ref{fig:Txy}(a) the amplitude of the matrix element $\left|T_{x+\ii y}\right|$ is never zero, being $9.7\times 10^{-3}$ its minimum value in the range plotted. Therefore, strictly speaking there is not a selection rule for $n$ but the transition is highly suppressed for $n>3$. In contrast with the former, Fig.~\ref{fig:Txy}(b) shows the existence of a selection rule for $\ket{1,1,1;1^+}\to\ket{n,1,1;2^+}$ so that the transition is only allowed for $\Delta n$ odd. Finally, from the mathematical expressions, it is clear that the transition is allowed for $\Delta m=0$, and as Fig.~\ref{fig:Iphi} suggests there are no restrictions over $\Delta l$.
 
\begin{widetext}
\begin{center}
\begin{tabular}{|c|c|c|c|}
\hline
{\bf Transition}  & $\Delta n$                   &  $\Delta m$                   & $\Delta l$                   \\
\hline
\hline
$T_z\left(\ket{1,1,1;1^+}\to\ket{n,m,l;1^+}\right)$  & \multirow{6}{*}{Odd}  & \multirow{6}{*}{Given by Eq.~(\ref{selectionrulesinM})}  & \multirow{6}{*}{0}  \\
$T_z\left(\ket{1,1,1;1^+}\to\ket{n,m,l;1^-}\right)$  &                     &                     &                     \\
$T_z\left(\ket{1,1,1;1^+}\to\ket{n,m,l;2^+}\right)$ &                     &                     &                     \\
$T_z\left(\ket{1,1,1;1^+}\to\ket{n,m,l;2^-}\right)$ &                     &                     &                     \\
$T_z\left(\ket{1,1,1;2^+}\to\ket{n,m,l;2^+}\right)$ &                     &                     &                     \\
$T_z\left(\ket{1,1,1;2^+}\to\ket{n,m,l;2^-}\right)$ &                     &                     &                     \\
\hline
$T_{x+\ii y}\left(\ket{1,1,1;1^+}\to\ket{n,m,l;1^+}\right)$ & \multirow{2}{*}{No restrictions (highly suppressed for $n>3$)} & \multirow{6}{*}{0} & \multirow{6}{*}{No restrictions} \\
$T_{x+\ii y}\left(\ket{1,1,1;1^+}\to\ket{n,m,l;1^-}\right)$ &                     &                     &                     \\
\cline{1-2}
$T_{x+\ii y}\left(\ket{1,1,1;1^+}\to\ket{n,m,l;2^+}\right)$ & \multirow{4}{*}{Odd} &                     &                     \\
$T_{x+\ii y}\left(\ket{1,1,1;1^+}\to\ket{n,m,l;2^-}\right)$ &                     &                     &                     \\
$T_{x+\ii y}\left(\ket{1,1,1;2^+}\to\ket{n,m,l;2^+}\right)$ &                     &                     &                     \\
$T_{x+\ii y}\left(\ket{1,1,1;2^+}\to\ket{n,m,l;2^-}\right)$ &                     &                     &                    \\
\hline
\end{tabular}
\captionof{table}{Selection rules for the transitions $T_\mu(\ket{1,1,1;q^s}\to\ket{n,m,l;q'^{s'}})$ with $n>1$. The values of $R$, $h$, and $\alpha$ are arbitrary.}
\label{tab:selectionrules}
\end{center}
\end{widetext}

\section{Conclusions}
In this article, we studied the effects of size and morphology
on the confined electronic states in TI-NPs, including the transition amplitudes and selection rules for transitions within the dipole approximation. Moreover, by means of a spatial representation of the probability density distribution, we identified the
conditions leading to the emergence of topologically trivial surface states arising from the underlying bulk band structure. The existence of such surface states is predicted in our theoretical formulation of the problem, since we incorporated a more general boundary condition involving the vanishing of the normal component of the probability current at the boundary~\cite{Johnson_75,Chodos_74A,Chodos_74B,Chodos_74C}, in contrast with the hard-wall applied in previous studies reported in the literature~\cite{Governale_20,Gioia_19}, where the full eigenstate vanishes at the boundary by construction. The predicted existence of these topologically trivial surface states arising form the bulk upon geometric confinement, supports the idea that they might couple and interact with the topologically protected surface states in TI nanostructures, as suggested by magnetotransport experiments~\cite{Tang_19,Liao_17,Tian_13,Steinberg_11,Zhao_13,Kim_12,Chiatti_16,Checkelsky_11}. The nature of the coupling mechanism, that in addition to confinement might involve disorder and many-body effects~\cite{Liao_17,Chiatti_16,Tian_13,Lu_11,Kim_12,Checkelsky_11}, is a subject of further study that goes beyond the scope of the present work.

\section*{Acknowledgements }

 J.D.C.-Y. and E.M. acknowledge financial support from ANID PIA Anillo ACT/192023. E.M. also acknowledges financial support from Fondecyt 1190361.

\section*{Author contributions statement}
E.M. conceived the study and supervised the work. J-D.C.-Y. and E.M. performed the analytical calculations. J.D.C.-Y. performed numerical calculations to generate all the figures. All authors reviewed the manuscript. 



\bibliography{BibPaper.bib}

\ifarXiv
    \foreach \x in {1,...,\numbersupplementpages}
    {
        \clearpage
        \includepdf[pages={\x,{}}]{\supplementfilename}
    }
\fi

\end{document}
\end{document}